\documentclass[journal,twoside]{IEEEtran}

\usepackage{graphicx}
\usepackage[T1]{fontenc}
\usepackage[bookmarks=false]{hyperref}

\usepackage{enumitem}
\usepackage[cmex10]{amsmath}
\usepackage{amssymb}
\usepackage{cite}
\usepackage{amsthm}
\usepackage{textcomp}
\usepackage{siunitx}
\usepackage{color}
\usepackage{subfigure}
\usepackage{cases}
\usepackage[font={small}]{caption}
\usepackage{bbm}

\newtheorem{theorem}{Theorem}
\newtheorem{lemma}{Lemma}

\theoremstyle{definition}
\newtheorem{example}{Example}
\newtheorem{definition}{Definition}

\theoremstyle{remark}
\newtheorem{remark}{Remark}
\newtheorem{claim}{Claim}

\setlist[description]{style=multiline}

\let\emptyset\varnothing

\IEEEoverridecommandlockouts

\begin{document}
\sloppy

\title{A Fundamental Tradeoff between Computation and Communication in Distributed Computing} 

\author{Songze~Li,~\IEEEmembership{Student~Member,~IEEE,}
        Mohammad~Ali~Maddah-Ali,~\IEEEmembership{Member,~IEEE,}
        Qian~Yu,~\IEEEmembership{Student~Member,~IEEE,}
        and~A.~Salman~Avestimehr,~\IEEEmembership{Senior Member,~IEEE}
\thanks{S.~Li, Q.~Yu and A.S.~Avestimehr are with the Department of Electrical Engineering, University of Southern California, Los Angeles, CA, 90089, USA (e-mail: songzeli@usc.edu;  qyu880@usc.edu; avestimehr@ee.usc.edu).}
\thanks{M. A. Maddah-Ali is with Department of Electrical Engineering, Sharif University of Technology, Tehran, 11365, Iran (e-mail: maddah\_ali@sharif.edu).}
\thanks{A preliminary part of this work was presented in 53rd Annual Allerton Conference on Communication, Control, and Computing, 2015~\cite{LMA_all}. A part of this work was presented in IEEE International Symposium on Information Theory, 2016~\cite{LMA_ISIT16}. A part of this work was presented in the 6th International Workshop on Parallel and Distributed Computing for Large Scale Machine Learning and Big Data Analytics, 2017~\cite{li2017coded}.}
\thanks{This work is in part supported by NSF grants CCF-1408639, NETS-1419632, ONR award N000141612189, NSA Award No. H98230-16-C-0255, and a research gift from Intel. This material is based upon work supported by Defense Advanced Research Projects Agency (DARPA) under Contract No. HR001117C0053. The views, opinions, and/or findings expressed are those of the author(s) and should not be interpreted as representing the official views or policies of the Department of Defense or the U.S. Government.}
}

\maketitle

\begin{abstract}
How can we optimally trade extra computing power to reduce the communication load in distributed computing? We answer this question by characterizing a fundamental tradeoff between computation and communication in distributed computing, i.e., the two are \emph{inversely proportional} to each other. 

More specifically, a general distributed computing framework, motivated by commonly used structures like MapReduce, is considered, where the overall computation is decomposed into computing a set of ``Map'' and ``Reduce'' functions distributedly across multiple computing nodes. A coded scheme, named ``Coded Distributed Computing'' (CDC), is proposed to demonstrate that increasing the computation load of the Map functions by a factor of $r$ (i.e., evaluating each function at $r$ \emph{carefully chosen} nodes) can create novel coding opportunities that reduce the communication load by the same factor.

An information-theoretic lower bound on the communication load is also provided, which matches the communication load achieved by the CDC scheme. As a result, the optimal computation-communication tradeoff in distributed computing is exactly characterized.

Finally, the coding techniques of CDC is applied to the Hadoop \texttt{TeraSort} benchmark to develop a novel \texttt{CodedTeraSort} algorithm, which is empirically demonstrated to speed up the overall job execution by $1.97\times$ - $3.39\times$, for typical settings of interest.

\end{abstract}


\section{Introduction}\label{sec:intro}
We consider a general distributed computing framework, motivated by prevalent structures like MapReduce~\cite{dean2004mapreduce} and Spark~\cite{zaharia2010spark}, in which the overall computation is decomposed into two stages: ``Map'' and ``Reduce''.   Firstly in the Map stage, distributed computing nodes process parts of the input data locally, generating some intermediate values according to their designed Map functions. Next, they exchange the calculated intermediate values among each other (a.k.a. data shuffling), in order to calculate the final output results distributedly using their designed Reduce functions.

Within this framework, data shuffling often appears to limit the performance of distributed computing applications, including self-join~\cite{ahmad2012tarazu}, tera-sort~\cite{guo2013ishuffle}, and machine learning algorithms~\cite{chowdhury2011managing}. For example, in a Facebook's Hadoop cluster, it is observed that 33\% of the overall job execution time is spent on data shuffling~\cite{chowdhury2011managing}. Also as is observed in~\cite{zhang2013performance}, 70\% of the overall job execution time is spent on data shuffling when running a self-join application on an Amazon EC2 cluster~\cite{amazonec2}. As such motivated, we ask this fundamental question that \emph{if coding can help distributed computing in reducing the load of communication and speeding up the overall computation?} Coding is known to be helpful in coping with the channel uncertainty in telecommunication and also in reducing the storage cost in distributed storage systems and cache networks. In this work, we extend the application of coding to \emph{distributed computing} and  propose a framework to substantially reduce the load of data shuffling via coding and some extra computing in the Map phase.

More specifically, we formulate and characterize a fundamental tradeoff relationship between ``computation load'' in the Map phase and ``communication load'' in the data shuffling phase, and demonstrate that the two are \emph{inversely proportional} to each other. We propose an optimal coded scheme, named ``Coded Distributed Computing'' (CDC), which demonstrates that increasing the computation load of the Map phase by a factor of $r$ (i.e., evaluating each Map function at $r$ \emph{carefully chosen} nodes) can create novel coding opportunities in the data shuffling phase that reduce the communication load by the same factor.

To illustrate our main result, consider a distributed computing framework to compute $Q$ arbitrary output functions from $N$ input files, using $K$ distributed computing nodes. As mentioned earlier, the overall computation is performed by computing a set of Map and Reduce functions distributedly across the $K$ nodes. In the Map phase, each input file is processed locally, in one of the nodes,  to generate $Q$ intermediate values, each corresponding to one of the $Q$ output functions. Thus, at the end of this phase, $QN$ intermediate values are calculated, which can be split into $Q$ subsets of $N$ intermediate values and each subset is needed to calculate one of the output functions. In the Shuffle phase, for every output function to be calculated, all $N$  intermediate values corresponding to that function are transferred to one of the nodes for reduction. Of course, depending on the node that has been chosen to reduce an output function, a part of the intermediate values are already available locally, and do not need to be transferred in the Shuffle phase. This is because that the Map phase has been carried out on the same set of nodes, and the results of mapping done at a node can remain in that node to be used for the Reduce phase.  This offers some saving in the load of communication. To reduce the communication load even more, we may map each input file in \emph{more than one} nodes.  Apparently, this increases the fraction of intermediate values that are locally available. However, as we will show, there is a better way to exploit this redundancy in computation to reduce the communication load.  The main message of this paper is to show that following a particular patten in repeating Map computations along with some coding techniques, we can significantly reduce the load of communication. Perhaps surprisingly, we show that the gain of coding in reducing communication load scales with the size of the network.

To be more precise, we define the \emph{computation load} $r$,  $1 \leq r\leq K$, as the total number of computed Map functions at the nodes, normalized by $N$. For example, $r=1$ means that  none  of the Map functions has been re-computed, and $r=2$ means that on average each Map function can be computed on two nodes. We also define \emph{communication load} $L$, $0 \leq L\leq 1$,  as the total amount of information exchanged across nodes in the shuffling phase, normalized by the size of $QN$ intermediate values, in order to compute the $Q$ output functions disjointly and uniformly across the $K$ nodes.  Based on this formulation, we now ask the following fundamental question:
\begin{itemize}[leftmargin=3mm]
\item \emph{Given a computation load $r$ in the Map phase, what is the minimum communication load $L^*(r)$, using any data shuffling scheme, needed to compute the final output functions?}
\end{itemize}

We propose Coded Distributed Computing (CDC) that achieves a communication load of $L_{\textup{coded}}(r) = \frac{1}{r}\cdot (1-\frac{r}{K})$ for $r=1,\ldots,K$, and the lower convex envelop of these points. CDC employs a specific strategy to assign the computations of the Map and Reduce functions across the computing nodes, in order to enable novel coding opportunities for data shuffling. In particular, for a computation load $r \in \{1,\ldots,K\}$, CDC utilizes a carefully designed repetitive mapping of data blocks at $r$ distinct nodes to create coded multicast messages that deliver data \emph{simultaneously} to a subset of $r \geq 1$ nodes.  Hence, compared with an uncoded data shuffling scheme, which as we show later achieves a communication load $L_{\textup{uncoded}}(r) = 1-\frac{r}{K}$, CDC is able to reduce the communication load by exactly a factor of the computation load $r$. Furthermore, the proposed CDC scheme applies to a more general distributed computing framework where every output function is computed by more than one, or particularly $s \in \{1,\ldots,K\}$ nodes, which provides better fault-tolerance in distributed computing.

\begin{figure}[htbp]
   \centering
   \includegraphics[width=0.4\textwidth]{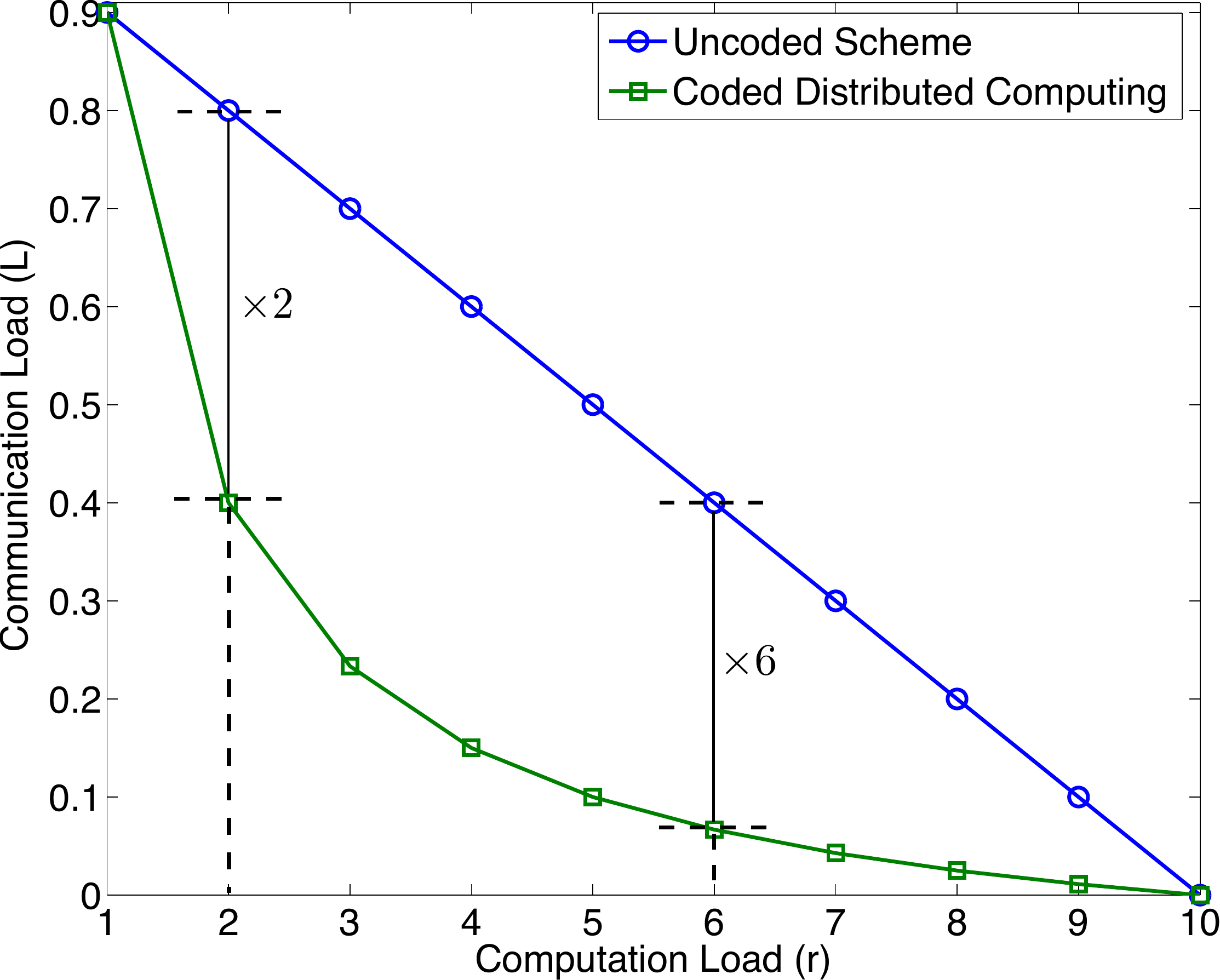}
   \caption{Comparison of the communication load achieved by Coded Distributed Computing $L_{\textup{coded}}(r)$ with that of the uncoded scheme $L_{\textup{uncoded}}(r)$, for $Q=10$ output functions, $N=2520$ input files and $K=10$ computing nodes. For $r \in \{1,\ldots, K\}$, CDC is $r$ times better than the uncoded scheme.}
   \label{fig:scaling}
\end{figure}

We numerically compare the computation-communication tradeoffs of CDC and uncoded data shuffling schemes (i.e., $L_{\textup{coded}}(r)$ and $L_{\textup{uncoded}}(r)$) in Fig.~\ref{fig:scaling}.  As it is illustrated, in the uncoded scheme that achieves a communication load $L_{\textup{uncoded}}(r) = 1-\frac{r}{K}$, increasing the computation load $r$ offers only a modest reduction in  communication load. In fact for any $r$, this gain vanishes for large number of nodes $K$.  Consequently,  it is not  justified to trade computation for communication using uncoded schemes. However, for the coded scheme that achieves a communication load of $L_{\textup{coded}}(r) = \frac{1}{r}\cdot (1-\frac{r}{K})$, increasing the computation load $r$ will significantly reduce the communication load, and this gain does not vanish for large $K$. For example as illustrated in Fig.~\ref{fig:scaling}, when mapping each file at one extra node ($r=2$), CDC reduces the communication load by 55.6\%, while the uncoded scheme only reduces it by 11.1\%.

We also prove an information-theoretic lower bound on the minimum communication load $L^*(r)$. To prove the lower bound, we derive a lower bound on the total number of bits communicated by any subset of nodes, using induction on the size of the subset. To derive the lower bound for a particular subset of nodes, we first establish a lower bound on the number of bits needed by one of the nodes to recover the intermediate values it needs to calculate its assigned output functions, and then utilize the bound on the number of bits communicated by the rest of the nodes in that subset, which is given by the inductive argument. The derived lower bound on $L^*(r)$ matches the communication load achieved by the CDC scheme for any computation load $1\leq r \leq K$. As a result, we \emph{exactly} characterize the optimal tradeoff between computation load and communication load in the following: $$ L^*(r) = L_{\textup{coded}}(r)= \frac{1}{r}  \cdot(1-\frac{r}{K}), \, r \in \{1,\ldots,K\}.$$ For general $1\leq r \leq K$, $L^*(r)$ is the lower convex envelop of the above points $\{(r,L_{\textup{coded}}(r)):r \in \{1,\ldots,K\}\}$. Note that for large $K$, $\frac{1}{r} \cdot (1-\frac{r}{K}) \approx \frac{1}{r}$, hence $L^*(r) \approx \frac{1}{r}$. This result reveals a fundamental inversely proportional relationship between computation load and communication load in distributed computing. This also illustrates that the gain of $\frac{1}{r}$ achieved by CDC is optimal and it cannot be improved by any other scheme (since $L_{\textup{coded}}(r)$ is an information-theoretic lower bound on $L^*(r)$ that applies to any data shuffling scheme).

Having theoretically characterized the optimal computation-communication tradeoff achieved by the proposed CDC scheme, we also empirically demonstrate the practical impact of this tradeoff. In particular,
we apply the coding techniques of CDC to a widely used Hadoop sorting benchmark \texttt{TeraSort}~\cite{TSpackage}, developing a novel coded distributed sorting algorithm \texttt{CodedTeraSort}~\cite{li2017coded}. We perform extensive experiments on Amazon EC2 clusters, and observe that for typical settings of interest, \texttt{CodedTeraSort} speeds up the overall execution of the conventional \texttt{TeraSort} by a factor of $1.97 \times$ - $3.39\times$.

Finally, we discuss some future directions to extend the results of this work. In particular, we consider topics including heterogeneous networks with asymmetric tasks, straggling/failing computing nodes, multi-stage computation tasks, multi-layer networks and structured topology, joint storage and computation optimization, and coded edge/fog computing.

\noindent \textbf{Related Works.} The problem of characterizing the minimum communication for distributed computing has been previously considered in several settings in both computer science and information theory communities. In~\cite{yao1979some}, a basic computing model is proposed, where two parities have $x$ and $y$ and aim to compute a boolean function $f(x,y)$ by exchanging the minimum number of bits between them. Also, the problem of minimizing the required communication for computing the modulo-two sum of distributed binary sources with symmetric  joint distribution was introduced in~\cite{korner1979encode}. Following these two seminal works, a wide range of communication problems in the scope of distributed computing have been studied (see, e.g., \cite{OE90,becker1998,kushilevitz2006,OR01,nazer2007computation,ramamoorthy2013}). The key differences distinguishing the setting in this paper from most of the prior ones are 1) We focus on the flow of communication in a general distributed computing framework, motivated by MapReduce, rather than the structures of the functions or the input distributions. 2) We do not impose any constraint on the numbers of output results, input data files and computing nodes (they can be arbitrarily large), 3) We do not assume any special property (e.g. linearity) of the computed functions.

The idea of efficiently creating and exploiting \emph{coded multicasting} was initially proposed in the context of cache networks in~\cite{maddah2014fundamental, maddah2013decentralized}, and extended in~\cite{ji2014fundamental, karamchandani2014hierarchical}, where caches pre-fetch part of the content in a way to enable coding during the content delivery, minimizing the network traffic. In this paper, we propose a framework to study the tradeoff between computation and communication in distributed computing. We demonstrate that the coded multicasting opportunities exploited in the above caching problems also exist in the data shuffling of distributed computing frameworks, which can be created by a strategy of repeating the computations of the Map functions specified by the Coded Distributed Computing (CDC) scheme. 

Finally, in a recent work~\cite{lee2015speeding},  the authors have proposed methods for utilizing codes to speed up some specific distributed machine learning algorithms. The considered problem in this paper differs from~\cite{lee2015speeding} in the following aspects. We propose a general methodology for utilizing coding in data shuffling that can be applied to any distributed computing framework with a MapReduce structure, regardless of the underlying application. In other words, any distributed computing algorithm that fits in the MapReduce framework can benefit from the proposed CDC solution. We also characterize the information-theoretic computation-communication tradeoff in such frameworks. Furthermore, the coding used in~\cite{lee2015speeding} is at the application layer (i.e., applying computation on coded data), while in this paper we focus on applying codes directly on the shuffled data. 

\section{Problem Formulation}\label{sec:def}
In this section, we formulate a general distributed computing framework motivated by MapReduce, and define the function characterizing the tradeoff between computation and communication.

We consider the problem of computing $Q$ arbitrary output functions from $N$ input files using a cluster of $K$ distributed computing nodes (servers), for some positive integers $Q,N,K \in \mathbb{N}$, with $N \geq K$. More specifically, given $N$ input files $w_1,\ldots,w_N \in \mathbb{F}_{2^F}$, for some $F \in \mathbb{N}$, the goal is to compute $Q$ output functions $\phi_1,\ldots,\phi_Q$, where $\phi_q:(\mathbb{F}_{2^F})^N \rightarrow \mathbb{F}_{2^B}$, $q \in \{1,\ldots,Q\}$ maps all input files to a length-$B$ binary stream $u_q = \phi_q(w_1,\ldots,w_N) \in \mathbb{F}_{2^B}$, for some $B \in \mathbb{N}$. 

\begin{figure}[htbp]
   \centering
   \includegraphics[width=0.48\textwidth]{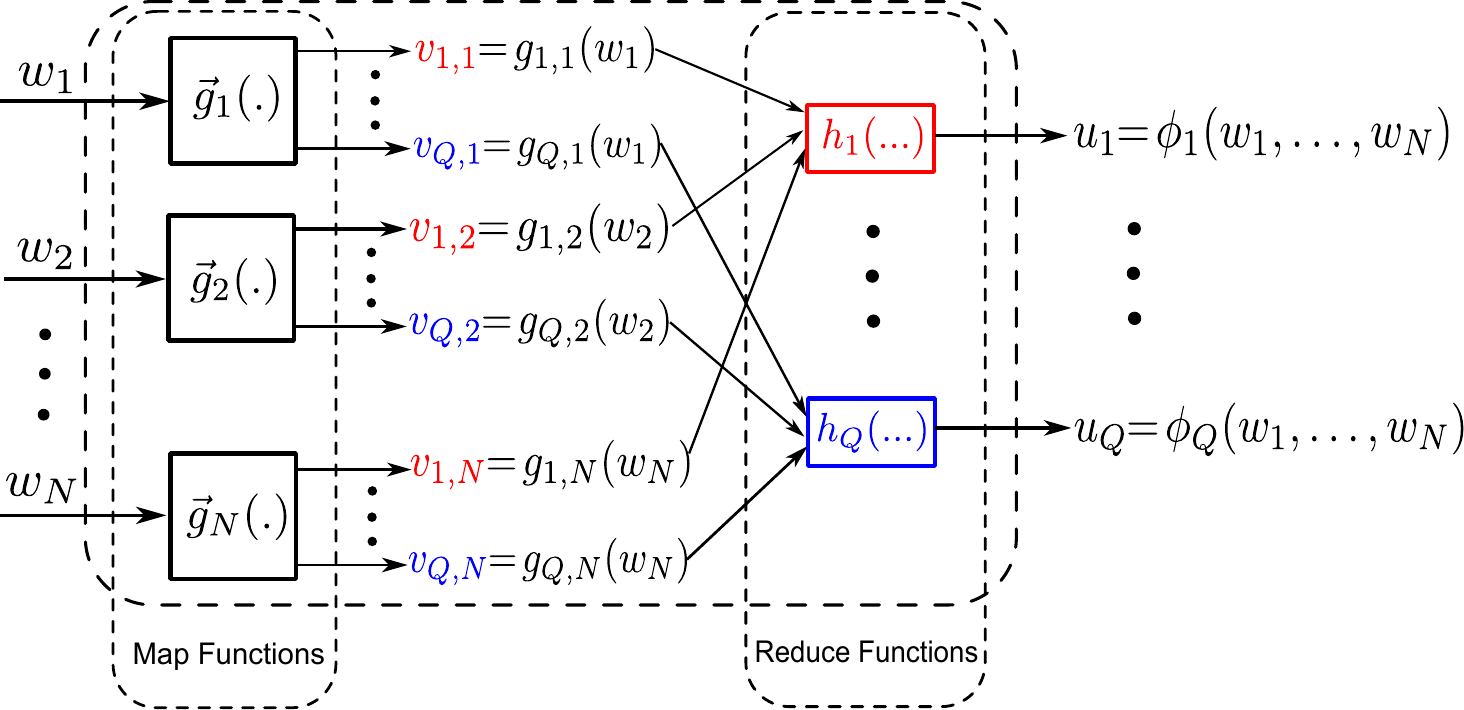}
   \caption{Illustration of a two-stage distributed computing framework. The overall computation is decomposed into computing a set of Map and Reduce functions.}
   \label{fig:frame}
\end{figure}

Motivated by MapReduce, we assume that as illustrated in Fig.~\ref{fig:frame} the computation of the output function $\phi_q$, $q \in \{1,\ldots,Q\}$ can be decomposed as follows:
\begin{equation}\label{eq:decom}
\phi_q(w_1,\ldots,w_N) = h_q(g_{q,1}(w_1),\ldots,g_{q,N}(w_N)),
\end{equation}
where
\begin{itemize}[leftmargin=4mm]
\item The ``Map'' functions $\vec{g}_{n} \!=\! (g_{1,n},\ldots,g_{Q,n})\!:\mathbb{F}_{2^F} \!\rightarrow \! (\mathbb{F}_{2^T})^Q$, $n \in \{1,\ldots,N\}$ maps the input file $w_n$ into $Q$ length-$T$ \emph{intermediate values} $v_{q,n}=g_{q,n}(w_n)\in  \mathbb{F}_{2^T}$, $q\in \{1,\ldots,Q\}$, for some $T \in \mathbb{N}$.\footnote{When mapping a file, we compute $Q$ intermediate values in parallel, one for each of the $Q$ output functions. The main reason to do this is that parallel processing can be efficiently performed for applications that fit into the MapReduce framework. In other words, mapping a file according to one function is only marginally more expensive than mapping according to all functions. For example, for the canonical Word Count job, while we are scanning a document to count the number of appearances of one word, we can simultaneously count the numbers of appearances of other words with marginally increased computation cost.}
\item The ``Reduce'' functions $h_{q}: (\mathbb{F}_{2^T})^N \!\rightarrow \! \mathbb{F}_{2^B}$, $q\in \{1,\ldots,Q\}$ maps the intermediate values of the output function $\phi_q$ in all input files into the output value $u_q=h_q(v_{q,1},\ldots,v_{q,N})$. 
\end{itemize}

\begin{remark}
Note that for every set of output functions $\phi_1,\ldots,\phi_Q$ such a Map-Reduce decomposition exists (e.g., setting $g_{q,n}$$'s$ to identity functions such that $g_{q,n}(w_n) = w_n$ for all $n=1,\ldots,N$, and $h_q$ to $\phi_q$ in (\ref{eq:decom})). However, such a decomposition is not unique, and in the distributed computing literature, there has been quite some work on developing appropriate decompositions of computations like join, sorting and matrix multiplication (see, e.g., \cite{dean2004mapreduce,rajaraman2011mining}), for them to be performed efficiently in a distributed manner. Here we do not impose any constraint on how the Map and Reduce functions are chosen (for example, they can be arbitrary linear or non-linear functions). $\hfill \square$
\end{remark}

The above computation is carried out by $K$ distributed computing nodes, labelled as Node~$1$$,\ldots, \,$Node~$K$. They are interconnected through a multicast network. Following the above decomposition, the computation proceeds in three phases: \emph{Map}, \emph{Shuffle} and \emph{Reduce}.

\noindent {\bf Map Phase:} 
Node $k$, $k\in\{1,\ldots,K\}$ computes the Map functions of a set of files $\mathcal{M}_k$, which are stored on Node $k$, for some design parameter $\mathcal{M}_k \subseteq \{w_1,\ldots,w_N\}$. For each file $w_n$ in $\mathcal{M}_k$, Node $k$ computes $\vec{g}_n(w_n)\!=\!(v_{1,n},\ldots,v_{Q,n})$. We assume that each file is mapped by at least one node, i.e., $\underset{k=1,\ldots,K}{\cup} \mathcal{M}_k=\{w_1,\ldots,w_N\}$. 

\begin{definition}[Computation Load]
We define the \emph{computation load}, denoted by $r$, $1 \leq r \leq K$, as the total number of Map functions computed across the $K$ nodes, normalized by the number of files $N$, i.e., $r \triangleq \frac{\sum_{k=1}^K |\mathcal{M}_k|}{N}$. The computation load $r$ can be interpreted as the average number of nodes that map each file. $\hfill \Diamond$
\end{definition}

\noindent {\bf Shuffle Phase:} 
Node $k$, $k\in\{1,\ldots,K\}$ is responsible for computing a subset of output functions, whose indices are denoted by a set $\mathcal{W}_k \subseteq \{1,\ldots,Q\}$. We focus on the case $\frac{Q}{K} \in \mathbb{N}$, and utilize a \emph{symmetric} task assignment across the $K$ nodes to maintain load balance. More precisely, we require 1) $|\mathcal{W}_1|=\cdots=|\mathcal{W}_K|=\frac{Q}{K}$, 2) $\mathcal{W}_j \cap \mathcal{W}_k=\emptyset$ for all $j \neq k$. 

\begin{remark}
Beyond the symmetric task assignment considered in this paper, characterizing the optimal computation-communication tradeoff allowing general asymmetric task assignments is a challenging open problem. As the first step to study this problem, in our follow-up work~\cite{QLMA-resource} in which the number of output functions $Q$ is fixed and the computing resources are abundant (e.g., number of computing nodes $K \gg Q$), we have shown that asymmetric task assignments can do better than the symmetric ones, and achieve the optimum run-time performance. $\hfill \square$
\end{remark}

To compute the output value $u_q$ for some $q \in \mathcal{W}_k$, Node $k$ needs the intermediate values that are \emph{not} computed \emph{locally} in the Map phase, i.e., $\{v_{q,n}: q \in \mathcal{W}_k, w_n \notin \mathcal{M}_k\}$. After Node $k$, $k \in \{1,\ldots,K\}$ has finished mapping all the files in ${\cal M}_k$, the $K$ nodes proceed to exchange the needed intermediate values. In particular, each node $k$, $k \in \{1,\ldots,K\}$, creates an input symbol $X_k \in \mathbb{F}_{2^{\ell_k}}$, for some $\ell_k \in \mathbb{N}$, as a function of the intermediate values computed locally during the Map phase, i.e., for some encoding function $\psi_k :(\mathbb{F}_{2^{T}})^{Q|{\cal M}_k|}\rightarrow \mathbb{F}_{2^{\ell_k}}$ at Node~$k$, we have
\begin{align}
X_k = \psi_k\left(\{\vec{g}_n:w_n \in \mathcal{M}_k\}\right).
\end{align}
Having generated the message $X_k$, Node~$k$ multicasts it to all other nodes.

By the end of the Shuffle phase, each of the $K$ nodes receives $X_1,\ldots,X_K$ free of error.

\begin{definition}[Communication Load]
We define the \emph{communication load}, denoted by $L$, $0 \leq L \leq 1$, as $L \triangleq \frac{\ell_1 +\cdots+\ell_K}{QNT}$. That is, $L$ represents the (normalized) total number of bits communicated by the $K$ nodes during the Shuffle phase.\footnote{For notational convenience, we define all variables in binary extension fields. However, one can consider arbitrary field sizes. For example, we can consider all intermediate values $v_{q,n}$, $q=1,\ldots,Q$, $n=1,\ldots,N$, to be in the field $\mathbb{F}_{p^T}$, for some prime number $p$ and positive integer $T$, and the symbol communicated by Node~$k$ (i.e., $X_k$), to be in the field $\mathbb{F}_{s^{\ell_k}}$ for some prime number $s$ and positive integer $\ell_k$, for all $k=1,\ldots,K$. In this case, the communication load can be defined as $L \triangleq \frac{(\ell_1 +\cdots+\ell_K) \log s}{QNT\log p}$.}  $\hfill\Diamond$
\end{definition}

\noindent {\bf Reduce Phase:} Node $k$, $k\in\{1,\ldots,K\}$, uses the messages $X_1,\ldots,X_K$ communicated in the Shuffle phase, and the local results from the Map phase $\{\vec{g}_n: w_n\in \mathcal{M}_k\}$ to construct inputs to the corresponding Reduce functions of $\mathcal{W}_k$, i.e., for each $q \in {\cal W}_k$ and some decoding function $\chi_k^q: \mathbb{F}_{2^{\ell_1}} \times \cdots \times \mathbb{F}_{2^{\ell_K}} \times (\mathbb{F}_{2^T})^{Q|{\cal M}_k|} \rightarrow (\mathbb{F}_{2^T})^N $, Node $k$ computes
\begin{align}
(v_{q,1},\ldots,v_{q,N}) = \chi_k^q\left(X_1,\ldots,X_K, \{\vec{g}_n: w_n\in \mathcal{M}_k\}\right).
\end{align}

Finally, Node $k$, $k\in\{1,\ldots,K\}$, computes the Reduce function $u_q=h_q(v_{q,1}\ldots v_{q,N})$ for all $q \in \mathcal{W}_k$.

We say that a computation-communication pair $(r,L) \in \mathbb{R}^2$ is \emph{feasible} if for any $\delta >0$ and sufficiently large $N$, there exist $\mathcal{M}_1,\ldots,\mathcal{M}_K$, $\mathcal{W}_1,\ldots,\mathcal{W}_K$, a set of encoding functions $\{\psi_k\}_{k=1}^K$, and a set of decoding functions $\{\chi_{k}^q: q \in {\cal W}_k\}_{k=1}^K$ that achieve a computation-communication pair $(\tilde{r},\tilde{L}) \in \mathbb{Q}^2$ such that $|r-\tilde{r}| \leq \delta$, $|L-\tilde{L}| \leq \delta$, and Node $k$ can successfully compute all the output functions whose indices are in $\mathcal{W}_k$, for all $k \in \{1,\ldots,K\}$.

\begin{definition}
We define the \emph{computation-communication function} of the distributed computing framework 
\begin{equation}
L^*(r) \triangleq \inf\{L: (r,L) \text{ is feasible}\}.  
\end{equation}
$L^*(r)$ characterizes the optimal tradeoff between computation and communication in this framework.     $\hfill\Diamond$
\end{definition}

\noindent {\it Example} (Uncoded Scheme). In the Shuffle phase of a simple ``uncoded'' scheme, each node receives the needed intermediate values sent uncodedly by some other nodes. Since a total of $QN$ intermediate values are needed across the $K$ nodes and $rN \cdot \frac{Q}{K}  = \frac{rQN}{K}$ of them are already available after the Map phase, the communication load achieved by the uncoded scheme 
\begin{equation}\label{eq:uncode}
L_{\text{uncoded}}(r) = 1-r/K. 
\end{equation}

\begin{remark}
After the Map phase, each node knows the intermediate values of all $Q$ output functions in the files it has mapped. Therefore, for a fixed file assignment and any symmetric assignment of the Reduce functions, specified by ${\cal W}_1,\ldots,{\cal W}_K$, we can satisfy the data requirements using the same data shuffling scheme up to relabelling the Reduce functions. In other words, the communication load is independent of the assignment of the Reduce functions.   $\hfill \square$
\end{remark}

In this paper, we also consider a generalization of the above framework, which we call ``cascaded distributed computing framework'', where after the Map phase, each Reduce function is computed by more than one, or particularly $s$ nodes, for some $s \in \{1,\ldots,K\}$. This generalized model is motivated by the fact that many distributed computing jobs require multiple rounds of Map and Reduce computations, where the Reduce results of the previous round serve as the inputs to the Map functions of the next round.  Computing each Reduce function at more than one node admits \emph{data redundancy} for the subsequent Map-function computations, which can help to improve the fault-tolerance and reduce the communication load of the next-round data shuffling. We focus on the case $\frac{Q}{{K \choose s}} \in \mathbb{N}$, and enforce a symmetric assignment of the Reduce tasks to maintain load balance. Particularly, we require that every subset of $s$ nodes compute a disjoint subset of $\frac{Q}{{K \choose s}}$ Reduce functions.

The feasible computation-communication triple $(r,s,L) \in \mathbb{R} \times \mathbb{N} \times \mathbb{R}$ is defined similar as before. We define the computation-communication function of the cascaded distributed computing framework
\begin{equation}
L^*(r,s) \triangleq \inf\{L: (r,s,L) \text{ is feasible}\}.  
\end{equation}  
    
\section{Main Results}
\begin{theorem}
The computation-communication function of the distributed computing framework, $L^*(r)$ is given by
\begin{align}
L^*(r) =L_{\textup{coded}}(r) \triangleq \tfrac{1}{r}\cdot (1-\tfrac{r}{K}), \quad r \in \{1,\ldots,K\}, \label{eq:minLoad} 
\end{align}
for sufficiently large $T$.
For general $1 \leq r \leq K$, $L^*(r)$ is the lower convex envelop of the above points $\{(r,\frac{1}{r}\cdot (1-\frac{r}{K})): r \in \{1,\ldots,K\}\}$.
\end{theorem}

We prove the achievability of Theorem~1 by proposing a \emph{coded} scheme, named Coded Distributed Computing, in Section~V. We demonstrate that no other scheme can achieve a communication load smaller than the lower convex envelop of the points $\{(r,\frac{1}{r}\cdot (1-\frac{r}{K})): r \in \{1,\ldots,K\}\}$ by proving the converse in Section~VI.

\begin{remark}
Theorem~1 exactly characterizes the optimal tradeoff between the computation load and the communication load in the considered distributed computing framework. $\hfill \square$
\end{remark}

\begin{remark}
For $r \in \{1,\ldots,K\}$, the communication load achieved in Theorem~1 is less than that of the uncoded scheme in (\ref{eq:uncode}) by a multiplicative factor of $r$, which equals the computation load and can grow unboundedly as the number of nodes $K$ increases if e.g. $r = \Theta(K)$. As illustrated in Fig.~\ref{fig:scaling} in Section~\ref{sec:intro}, while the communication load of the uncoded scheme decreases linearly as the computation load increases, $L_{\textup{coded}}(r)$ achieved in Theorem~1 is inversely proportional to the computation load.$\hfill \square$
\end{remark} 

\begin{remark}
While increasing the computation load $r$ causes a longer Map phase, the coded achievable scheme of Theorem~1 maximizes the reduction of the communication load using the extra computations. Therefore, Theorem~1 provides an analytical framework to optimally trading the computation power in the Map phase for more bandwidth in the Shuffle phase, which helps to minimize the overall execution time of applications whose performances are limited by data shuffling.
$\hfill \square$
\end{remark} 

\begin{theorem}
The computation-communication function of the cascaded distributed computing framework, $L^*(r,s)$, for $r \in \{1,\ldots,K\}$, is characterized by
\begin{align}{\label{eq:cascade}}
L^*(r,s) = L_{\textup{coded}}(r,s) \triangleq \sum \limits_{\ell = \max\{r+1,s\}}^{\min \{r+s,K\}} \!\! \frac{\ell {K \choose \ell} {\ell -2 \choose r -1} {r \choose \ell -s}}{r {K \choose r}{K \choose s}},
\end{align}
for some $s \in \{1,\ldots,K\}$ and sufficiently large $T$. For general $1 \leq r \leq K$, $L^*(r,s)$ is the lower convex envelop of the above points $\{(r,L_{\textup{coded}}(r,s)): r \in \{1,\ldots,K\}\}$.
\end{theorem}

We present the Coded Distributed Computing scheme that achieves the computation-communication function in Theorem~2 in Section~V, and the converse of Theorem~2 in Section~VII.

\begin{remark}
A preliminary part of this result, in particular the achievability for the special case of $s\!=\!1$, or the achievable scheme of Theorem~1 was presented in~\cite{LMA_all}. We note that when $s=1$, Theorem~2 provides the same result as in Theorem~1, i.e., $L^*(r,1) = \frac{1}{r}\cdot (1-\frac{r}{K})$, for $r \in \{1,\ldots,K\}$. $\hfill \square$ 
\end{remark}

\begin{figure}[htbp]
   \centering
   \includegraphics[width=0.42\textwidth]{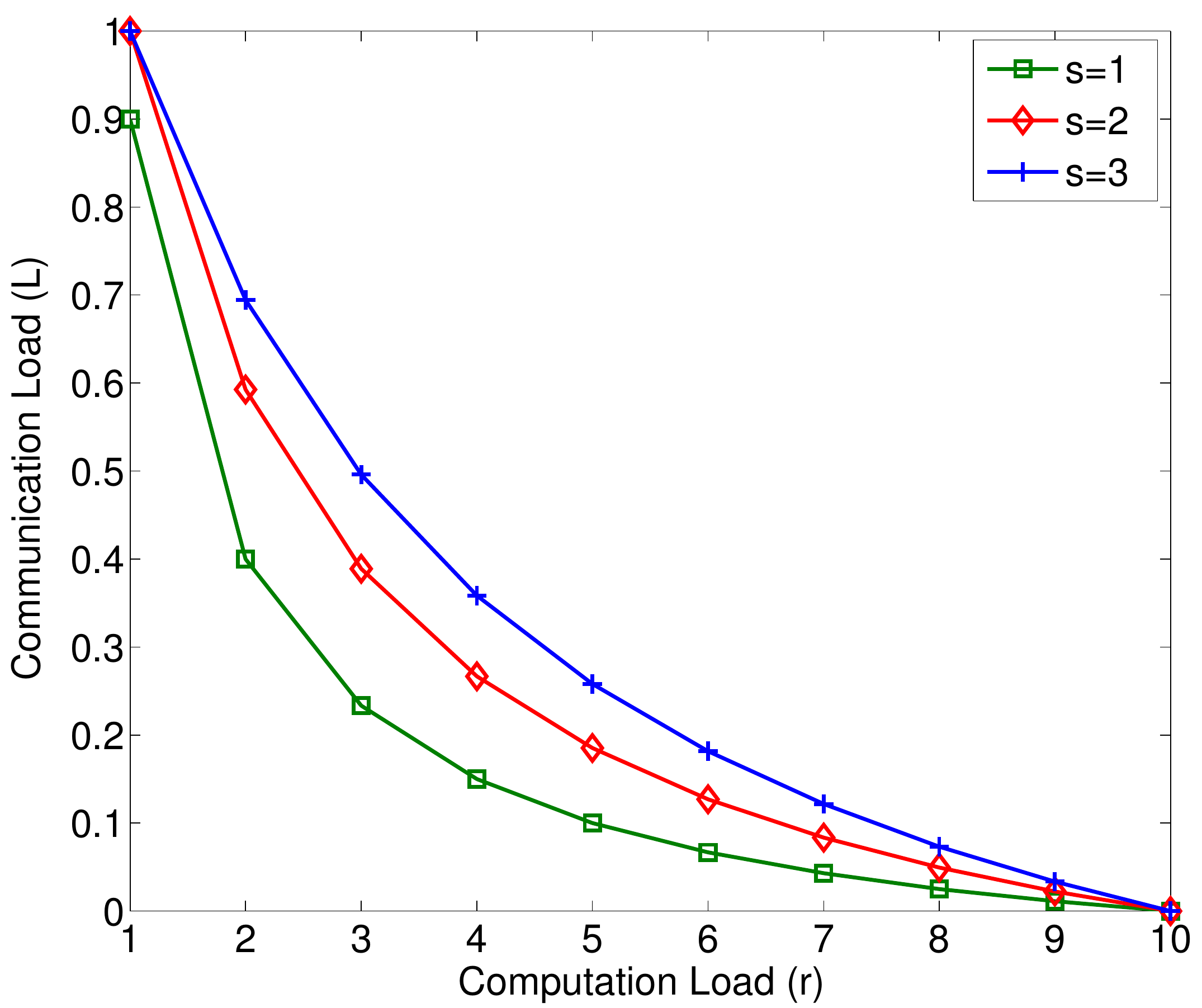}
   \caption{Minimum communication load $L^*(r,s)=L_{\textup{coded}}(r,s)$ in Theorem~2, for $Q\!=\!360$ output functions, $N\!=\!2520$ input files and $K\!=\!10$ computing nodes.}
   \label{fig:cascade}
\end{figure}

\begin{remark}
For any fixed $s \in \{1,\ldots,K\}$ (number of nodes that compute each Reduce function), as illustrated in Fig.~\ref{fig:cascade}, the communication load achieved in Theorem~2 outperforms the linear relationship between computation and communication, i.e., it is superlinear with respect to the computation load $r$. $\hfill \square$
\end{remark}

Before we proceed to describe the general achievability scheme for the cascaded distributed computing framework (also the distributed computing framework as a special case of $s=1$), we first illustrate the key ideas of the proposed Coded Distributed Computing scheme by presenting two examples in the next section, for the cases of $s=1$ and $s>1$ respectively. 

\section{Illustrative Examples: Coded Distributed Computing}\label{sec:example}
In this section, we present two illustrative examples of the proposed achievable scheme for Theorem 1 and Theorem 2, which we call Coded Distributed Computing (CDC), for the cases of $s=1$ (Theorem~1) and $s>1$ (Theorem~2) respectively.

\begin{example}[CDC for $s=1$]
We consider a MapReduce-type problem in Fig.~\ref{fig:MR} for distributed computing of $Q=3$ output functions, represented by red/circle, green/square, and blue/triangle respectively, from $N=6$ input files, using $K=3$ computing nodes. Nodes~$1$, $2$, and $3$ are respectively responsible for final reduction of red/circle, green/square, and  blue/triangle output functions. Let us first consider the case where no redundancy is imposed on the computations, i.e., each file is mapped once and computation load $r=1$. As shown in Fig.~\ref{fig:UncodedMR}, Node $k$ maps File $2k-1$ and File $2k$ for $k=1,2,3$. In this case, each node maps $2$ input files locally, computing all three intermediate values needed for the three output functions from each mapped file. In Fig.~\ref{fig:MR}, we represent, for example, the intermediate value of the red/circle function in File $n$ using a red circle labelled by $n$, for all $n=1,\ldots,6$. Similar representations follow for the green/square and the blue/triangle functions. After the Map phase, each node obtains $2$ out of $6$ required intermediate values to reduce the output function it is responsible for (e.g., Node~1 knows the red circles in File 1 and File 2). Hence, each node needs $4$ intermediate values  from the other nodes, yielding a communication load of $\frac{4 \times 3}{3 \times 6}=\frac{2}{3}$.

\begin{figure}[htbp]
  \centering
  \subfigure[Uncoded Distributed Computing Scheme.]{\includegraphics[width=0.42\textwidth, trim = 0cm 0cm 0cm 0cm]{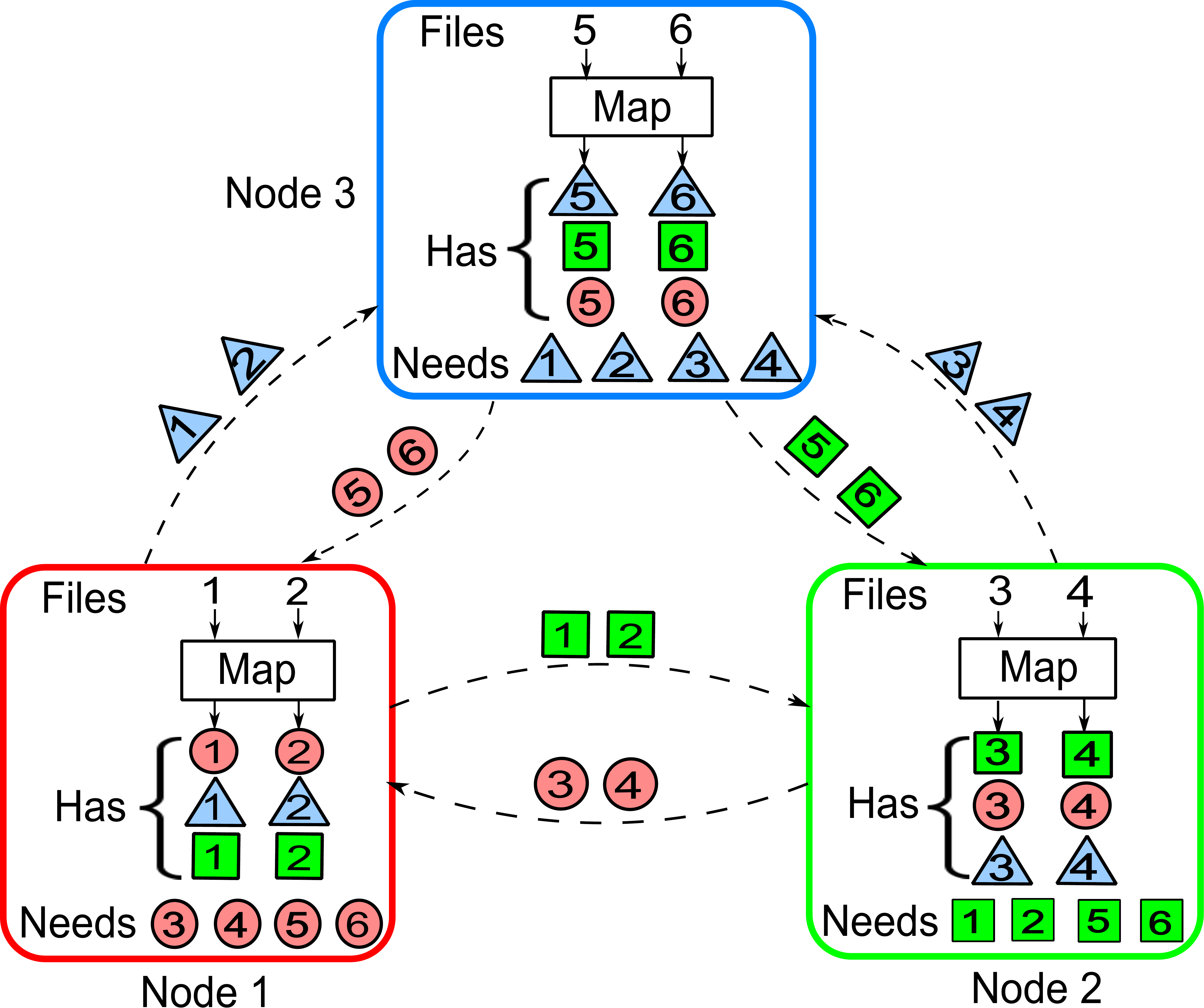}
      \label{fig:UncodedMR}}
      \hfill
      \subfigure[Coded Distributed Computing Scheme.]{\includegraphics[width=0.48\textwidth]{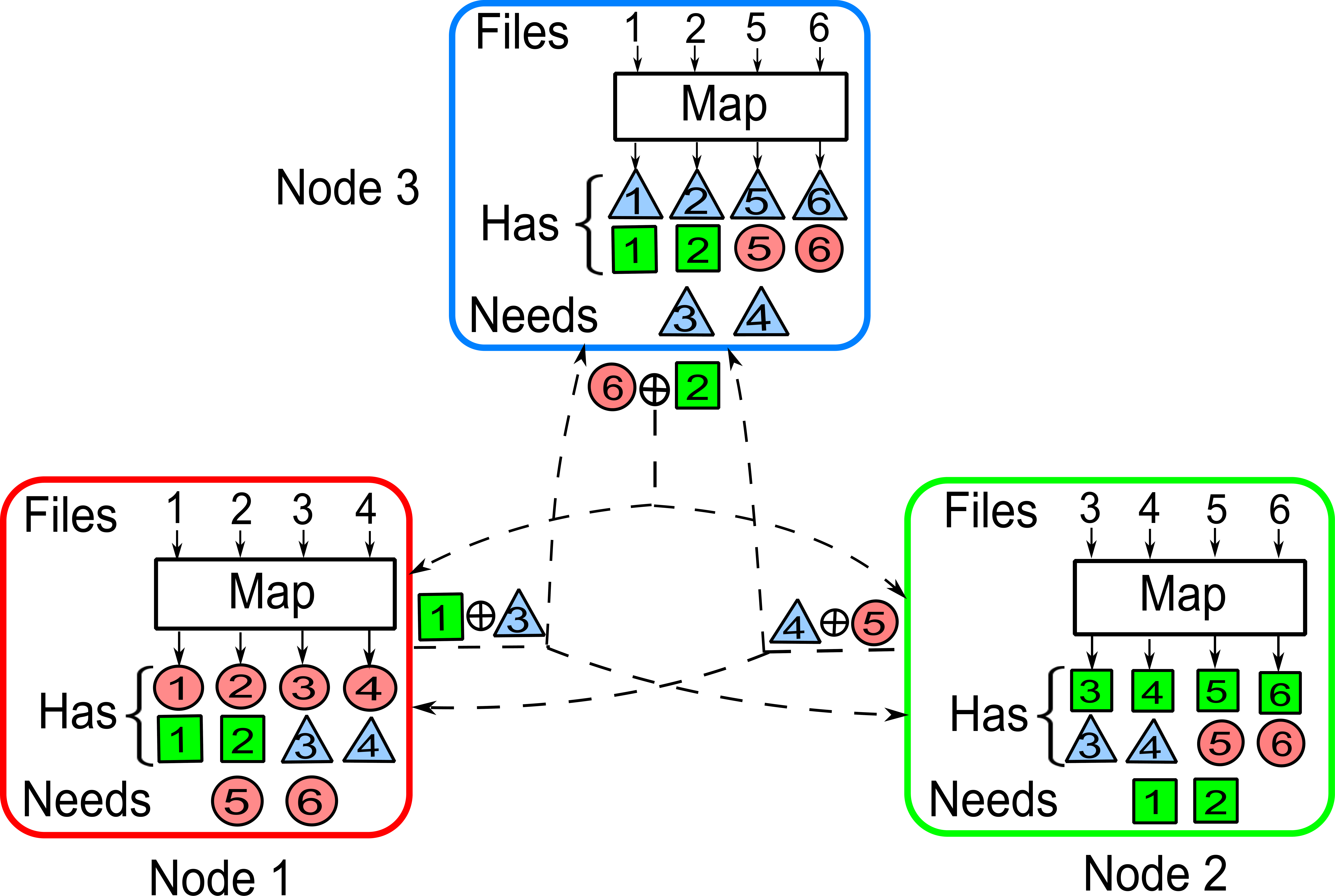}
      \label{fig:CodedMR}}
  \caption{Illustrations of the conventional uncoded distributed computing scheme with computation load $r=1$, and the proposed Coded Distributed Computing scheme with computation load $r=2$, for computing $Q=3$ functions from $N=6$ inputs on $K=3$ nodes.}
  \label{fig:MR}
\end{figure}

Now, we demonstrate how the proposed CDC scheme  trades the computation load to slash the communication load via in-network coding. As shown in Fig.~\ref{fig:CodedMR}, we double the computation load such that each file is now mapped on two nodes ($r=2$). It is apparent that since more local computations are performed, each node now only requires $2$ other intermediate values, and an uncoded shuffling scheme would achieve a communication load of $\frac{2 \times 3}{3 \times 6}=\frac{1}{3}$. However, we can do much better with coding. As shown in Fig.~\ref{fig:CodedMR}, instead of unicasting individual intermediate values, every node multicasts a bit-wise XOR, denoted by $\oplus$, of $2$ locally computed intermediate values to the other two nodes, simultaneously satisfying their data demands. For example, knowing the blue/triangle in File~$3$, Node~$2$ can cancel it from the coded packet sent by Node~$1$, recovering the needed green/square in File~$1$. Therefore, this coding incurs a communication load of $\frac{3}{3 \times 6}=\frac{1}{6}$, achieving a $2\times$ gain from the uncoded shuffling.
$\hfill$ $\square$
\end{example}

From the above example, we see that for the case of $s=1$, i.e., each of the $Q$ output functions is computed on one node and the computations of the Reduce functions are symmetrically distributed across nodes, the proposed CDC scheme only requires performing bit-wise XOR as the encoding and decoding operations. However, for the case of $s>1$, as we will show in the following example, the proposed CDC scheme requires computing linear combinations of the intermediate values during the encoding process.

\begin{example}[CDC for $s>1$]
In this example, we consider a job of computing $Q=6$ output functions from $N=6$ input files, using $K=4$ nodes. We focus on the case where the computation load $r = 2$, and each Reduce function is computed by $s = 2$ nodes. In the Map phase, each file is mapped by $r=2$ nodes. As shown in Fig.~\ref{fig:CDC}, the sets of the files mapped by the $4$ nodes are $\mathcal{M}_1=\{w_1,w_2,w_3\}$, $\mathcal{M}_2=\{w_1,w_4,w_5\}$, $\mathcal{M}_3=\{w_2,w_4,w_6 \}$, and $\mathcal{M}_4=\{w_3,w_5,w_6\}$. After the Map phase, Node $k$, $k \in \{1,2,3,4\}$, knows the intermediate values of all $Q=6$ output functions in the files in $\mathcal{M}_k$, i.e., $\{v_{q,n}:q \in \{1,\ldots,6\}, w_n \in \mathcal{M}_k\}$.  In the Reduce phase, we assign the computations of the Reduce functions in a symmetric manner such that every subset of  $s\!=\!2$ nodes compute a common Reduce function. More specifically as shown in Fig.~\ref{fig:CDC}, the sets of indices of the Reduce functions computed by the $4$ nodes are $\mathcal{W}_1\!=\!\{1,2,3\}$, $\mathcal{W}_2\!=\!\{1,4,5\}$, $\mathcal{W}_3\!=\!\{2,4,6 \}$, and $\mathcal{W}_4\!=\!\{3,5,6\}$. Therefore, for example, Node~1 still needs the intermediate values $\{v_{q,n}\!:\! q\in\{1,2,3\}, n \in \{4,5,6\}\}$ through data shuffling to compute its assigned Reduce functions $h_1$, $h_2$, $h_3$.

\begin{figure}[htbp]
   \centering
   \includegraphics[width=0.48\textwidth]{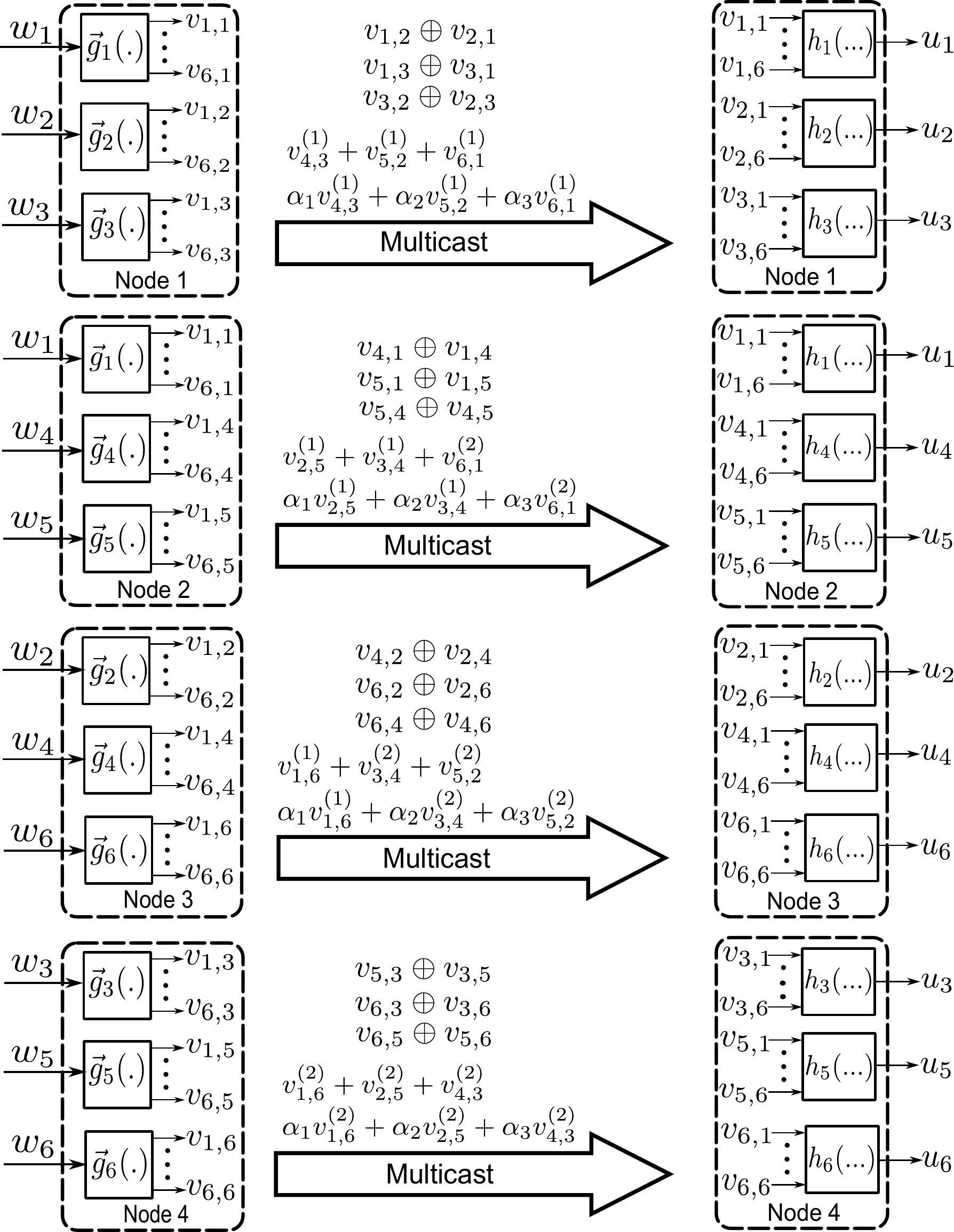}
   \caption{Illustration of the CDC scheme to compute $Q=6$ output functions from $N=6$ input files distributedly at $K=4$ computing nodes. Each file is mapped by $r=2$ nodes and each output function is computed by $s=2$ nodes. After the Map phase, every node knows $6$ intermediate values, one for each output function, in every file it has mapped. The Shuffle phase proceeds in two rounds. In the first round, each node multicasts bit-wise XOR of intermediate values to subsets of two nodes. In the second round, each node splits an intermediate value $v_{q,n}$ evenly into two segments $v_{q,n}=(v_{q,n}^{(1)},v_{q,n}^{(2)})$, and multicasts two linear combinations of the segments that are cosntructed using coefficients $\alpha_1$, $\alpha_2$, and $\alpha_3$ to the other three nodes.}
   \label{fig:CDC}
\end{figure}

The data shuffling process consists of two rounds of communication over the multicast network. In the first round, intermediate values are communicated within each subset of $3$ nodes. In the second round, intermediate values are communicated within the set of all $4$ nodes. In what follows, we describe these two rounds of communication respectively.

\noindent \emph{Round 1: Subsets of $3$ nodes.} We first consider the subset $\{1,2,3\}$. During the data shuffling, each node whose index is in $\{1,2,3\}$ multicasts a bit-wise XOR of two locally computed intermediate values to the other two nodes: \begin{itemize}
\item Node 1 multicasts $v_{1,2} \oplus v_{2,1}$ to Node~$2$ and Node~$3$,
\item Node 2 multicasts $v_{4,1} \oplus v_{1,4}$ to Node~$1$ and Node~$3$,
\item Node 3 mulicasts $v_{4,2} \oplus v_{2,4}$ to Node~$1$ and Node~$2$,
\end{itemize}

Since Node 2 knows $v_{2,1}$ and Node 3 knows $v_{1,2}$ locally, they can respectively decode $v_{1,2}$ and $v_{2,1}$ from the coded message $v_{1,2} \oplus v_{2,1}$.

We employ the similar coded shuffling scheme on the other 3 subsets of 3 nodes. After the first round of shuffling,
\begin{itemize}
\item Node 1 recovers $(v_{1,4}, v_{1,5})$, $(v_{2,4},v_{2,6})$ and $(v_{3,5},v_{3,6})$,
\item Node 2 recovers $(v_{1,2}, v_{1,3})$, $(v_{4,2},v_{4,6})$ and $(v_{5,3},v_{5,6})$,
\item Node 3 recovers $(v_{2,1}, v_{2,3})$, $(v_{4,1},v_{4,5})$ and $(v_{6,3},v_{6,5})$,
\item Node 4 recovers $(v_{3,1}, v_{3,2})$, $(v_{5,1},v_{5,4})$ and $(v_{6,2},v_{6,4})$.
\end{itemize}

\noindent \emph{Round 2: All $4$ nodes.} We first split each of the intermediate values $v_{6,1}$, $v_{5,2}$, $v_{4,3}$, $v_{3,4}$, $v_{2,5}$, and $v_{1,6}$ into two equal-sized segments each containing $T/2$ bits, which are denoted by $v_{q,n}^{(1)}$ and $v_{q,n}^{(2)}$ for an intermediate value $v_{q,n}$. Then, for some coefficients $\alpha_1,\alpha_2, \alpha_3 \in \mathbb{F}_{2^{\frac{T}{2}}}$, Node~$1$ multicasts the following two linear combinations of three locally computed segments to the other three nodes.
\begin{align}
&v_{4,3}^{(1)} + v_{5,2}^{(1)} +  v_{6,1}^{(1)},\\
&\alpha_1 v_{4,3}^{(1)} + \alpha_2 v_{5,2}^{(1)} + \alpha_3 v_{6,1}^{(1)}.
\end{align}

Similarly, as shown in Fig.~\ref{fig:CDC}, each of Node~$2$, Node~$3$, and Node~$4$ multicasts two linear combinations of three locally computed segments to the other three nodes, using the same coefficients $\alpha_1$, $\alpha_2$, and $\alpha_3$.

Having received the above two linear combinations, each of Node~$2$, Node~$3$, and Node~$4$ first subtracts out one segment available locally from the combinations, or more specifically, $v_{6,1}^{(1)}$ for Node~$2$, $v_{5,2}^{(1)}$ for Node~$3$, and $v_{4,3}^{(1)}$ for Node~$4$. After the subtraction, each of these three nodes recovers the required segments from the two linear combinations. More specifically, Node 2 recovers $v_{4,3}^{(1)}$ and $v_{5,2}^{(1)}$, Node 3 recovers $v_{4,3}^{(1)}$ and $v_{6,1}^{(1)}$, and Node 4 recovers $v_{5,2}^{(1)}$ and $v_{6,1}^{(1)}$. It is not difficult to see that the above decoding process is guaranteed to be successful if $\alpha_1$, $\alpha_2$, and $\alpha_3$ are all distinct from each other, which requires the field size $2^{\frac{T}{2}} \geq 3$ (e.g., $T=4$). Following the similar procedure, each node recovers the required segments from the linear combinations multicast by the other three nodes. More specifically, after the second round of data shuffling,
\begin{itemize}
\item Node 1 recovers $v_{1,6}$, $v_{2,5}$ and $v_{3,4}$,
\item Node 2 recovers $v_{1,6}$, $v_{4,3}$ and $v_{5,2}$,
\item Node 3 recovers $v_{2,5}$, $v_{4,3}$ and $v_{6,1}$,
\item Node 4 recovers $v_{3,4}$, $v_{5,2}$ and $v_{6,1}$.
\end{itemize}

We finally note that in the second round of data shuffling, each linear combination multicast by a node is simultaneously useful for the rest of the three nodes.
$\hfill \square$
\end{example}

\section{General Achievable Scheme: Coded Distributed Computing}\label{sec:scheme}
In this section, we formally prove the upper bounds in Theorem 1 and 2 by presenting and analyzing the Coded Distributed Computing (CDC) scheme. We focus on the more general case considered in Theorem 2 with $s \geq 1$, and the scheme for Theorem 1 simply follows by setting $s=1$. 

We first consider the integer-valued computation load $r \in \{1,\ldots,K\}$, and then generalize the CDC scheme for any $1\leq r \leq K$. When $r=K$, every node can map all the input files and compute all the output functions locally, thus no communication is needed and $L^*(K,s)=0$ for all $s \in \{1,\ldots,K\}$. In what follows, we focus on the case where $r < K$.

We consider sufficiently large number of input files $N$, and ${K \choose r}(\eta_1-1) < N \leq {K \choose r}\eta_1$, for some $\eta_1 \in \mathbb{N}$. We first inject ${K \choose r}\eta_1-N$ empty files into the system to obtain a total of $\bar{N} = {K \choose r}\eta_1$ files, which is now a multiple of of ${K \choose r}$. We note that $\lim \limits_{N \rightarrow \infty} \frac{\bar{N}}{N}=1$. Next, we proceed to present the achievable scheme for a system with $\bar{N}$ input files $w_1,\ldots,w_{\bar{N}}$.

\subsection{Map Phase Design}\label{sec:map}
In the Map phase the $\bar{N}$ input files are evenly partitioned into ${K \choose r}$ disjoint batches of size $\eta_1$, each corresponding to a subset $\mathcal{T} \subset \{1,\ldots,K\}$ of size $r$, i.e.,
\begin{equation}
\{w_1,\ldots,w_{\bar{N}}\} = \underset{{\cal T} \subset \{1,\ldots,K\}, |{\cal T}|=r}{\cup}\mathcal{B}_{\cal T},
\end{equation}
where $\mathcal{B}_{\cal T}$ denotes the batch of $\eta_1$ files corresponding to the subset $\mathcal{T}$.

Given this partition, Node $k$, $k \in \{1,\ldots,K\}$, computes the Map functions of the files in $\mathcal{B}_{\cal T}$ if $k \in \mathcal{T}$. Or equivalently, $\mathcal{B}_{\cal T} \subseteq \mathcal{M}_k$ if $k \in \mathcal{T}$. Since each node is in ${K-1 \choose r-1}$ subsets of size $r$, each node computes ${K-1 \choose r-1}\eta_1=\frac{r\bar{N}}{K}$ Map functions, i.e., $|\mathcal{M}_k|=\frac{r\bar{N}}{K}$ for all $k \in \{1,\ldots,K\}$. After the Map phase, Node $k$, $k \in \{1,\ldots,K\}$, knows the intermediate values of all $Q$ output functions in the files in $\mathcal{M}_k$, i.e., $\{v_{q,n}:q \in \{1,\ldots,Q\}, w_n \in \mathcal{M}_k\}$.

\subsection{Coded Data Shuffling}\label{sec:shuffle}
We recall that we focus on the case where the number of the output functions $Q$ satisfies $\frac{Q}{{K \choose s}} \in \mathbb{N}$, and enforce a symmetric assignment of the Reduce functions such that every subset of $s$ nodes reduce $\frac{Q}{{K \choose s}}$ functions. Specifically, $Q = {K \choose s}\eta_2$ for some $\eta_2 \in \mathbb{N}$, and the computations of the Reduce functions are assigned symmetrically across the $K$ nodes as follows. Firstly the $Q$ Reduce functions are evenly partitioned into ${K \choose s}$ disjoint batches of size $\eta_2$, each corresponding to a unique subset $\mathcal{P}$ of $s$ nodes, i.e.,
\begin{equation}
\{1,\ldots,Q\} = \underset{{\cal P} \subseteq \{1,\ldots,K\}, |{\cal P}|=s }{\cup}\mathcal{D}_{\cal P},
\end{equation}
where $\mathcal{D}_{\cal P}$ denotes the indices of the batch of $\eta_2$ Reduce functions corresponding to the subset $\mathcal{P}$.

Given this partition, Node $k$, $k \in \{1,\ldots,K\}$, computes the Reduce functions whose indices are in $\mathcal{D}_{\cal P}$ if $k \in \mathcal{P}$. Or equivalently, $\mathcal{D}_{\cal P} \subseteq \mathcal{W}_k$ if $k \in \mathcal{P}$. As a result, each node computes ${K-1 \choose s-1}\eta_2=\frac{sQ}{K}$ Reduce functions, i.e., $|\mathcal{W}_k|=\frac{sQ}{K}$ for all $k \in \{1,\ldots,K\}$. 

For a subset ${\cal S}$ of $\{1,\ldots,K\}$ and ${\cal S}_1 \subset \mathcal{S}$ with $|{\cal S}_1| = r$, we denote the set of intermediate values needed by \emph{all} nodes in ${\cal S} \backslash\mathcal{S}_1$, \emph{no} node outside $\mathcal{S}$, and known \emph{exclusively} by nodes in $\mathcal{S}_1$ as $\mathcal{V}_{\mathcal{S}_1}^{\mathcal{S} \backslash \mathcal{S}_1}$. More formally:
\begin{align}
\mathcal{V}_{\mathcal{S}_1}^{\mathcal{S} \backslash \mathcal{S}_1} \triangleq \{v_{q,n}: &q \in \underset{k \in \mathcal{S} \backslash \mathcal{S}_1}{\cap} {\cal W}_k, q\notin \underset{k \notin \mathcal{S}}{\cup} {\cal W}_k, \nonumber\\
&w_n \in \underset{k \in {\cal S}_1}{\cap} \mathcal{M}_k, w_n\notin \underset{k \notin \mathcal{S}_1}{\cup} {\cal M}_k\}.\label{eq:V}
\end{align}

We observe that the set $\mathcal{V}_{\mathcal{S}_1}^{{\cal S} \backslash \mathcal{S}_1}$ defined above contains intermediate values of ${r\choose |{\cal S}| -s }\eta_2$ output functions. This is because that the output functions whose intermediate values are included in $\mathcal{V}_{\mathcal{S}_1}^{{\cal S} \backslash \mathcal{S}_1}$ should be computed \emph{exclusively} by the nodes in ${\cal S} \backslash \mathcal{S}_1$ and a subset of $s-(|{\cal S}|-r)$ nodes in ${\cal S}_1$. Therefore, $\mathcal{V}_{\mathcal{S}_1}^{{\cal S} \backslash \mathcal{S}_1}$ contains the intermediate values of a total of ${r\choose s - (|{\cal S}| -r)}\eta_2 = {r\choose |{\cal S}| -s }\eta_2$ output functions. Since every subset of $r$ nodes map a unique batch of $\eta_1$ files, $\mathcal{V}_{\mathcal{S}_1}^{{\cal S} \backslash \mathcal{S}_1}$ contains  $|\mathcal{V}_{\mathcal{S}_1}^{{\cal S} \backslash \mathcal{S}_1}| = {r\choose |{\cal S}| -s}\eta_1\eta_2$ intermediate values. 

Next, we first concatenate all intermediate values in $\mathcal{V}_{\mathcal{S}_1}^{{\cal S} \backslash \mathcal{S}_1}$ to construct a symbol $U_{\mathcal{S}_1}^{{\cal S} \backslash \mathcal{S}_1} \in \mathbb{F}_{2^{{r\choose |{\cal S}| -s}\eta_1\eta_2T}}$. Then for ${\cal S}_1 = \{\sigma_1,\ldots,\sigma_r\}$, we arbitrarily and evenly split $U_{\mathcal{S}_1}^{{\cal S} \backslash \mathcal{S}_1}$ into $r$ segments, each containing ${r \choose |{\cal S}| -s}\frac{\eta_1\eta_2T}{r}$ bits, i.e.,
\begin{align}
U_{\mathcal{S}_1}^{{\cal S} \backslash \mathcal{S}_1} = \left(U_{\mathcal{S}_1,\sigma_1}^{{\cal S} \backslash \mathcal{S}_1},U_{\mathcal{S}_1,\sigma_2}^{{\cal S} \backslash \mathcal{S}_1},\ldots,U_{\mathcal{S}_1,\sigma_r}^{{\cal S} \backslash \mathcal{S}_1}\right),
\end{align}
where $U_{\mathcal{S}_1,\sigma_i}^{{\cal S} \backslash \mathcal{S}_1} \in \mathbb{F}_{2^{{r\choose |{\cal S}| -s}\frac{\eta_1\eta_2T}{r}}}$ denotes the segment associated with Node~$\sigma_i \in {\cal S}_1$.

For each $k \in {\cal S}$, there are a total of ${|{\cal S}|-1 \choose r-1}$ subsets of ${\cal S}$ with size $r$ that contain the element $k$. We index these subsets as ${\cal S}_{(k)}[1],{\cal S}_{(k)}[2]\ldots,{\cal S}_{(k)}[{|{\cal S}|-1 \choose r-1}]$. Within a subset ${\cal S}_{(k)}[i]$, the segment associated with Node $k$ is $U_{\mathcal{S}_{(k)}[i],k}^{{\cal S} \backslash \mathcal{S}_{(k)}[i]}$, for all $i=1,\ldots,{|{\cal S}|-1 \choose r-1}$. We note that each segment $U_{\mathcal{S}_{(k)}[i],k}^{{\cal S} \backslash \mathcal{S}_{(k)}[i]}$, $i =  1,\ldots,{|{\cal S}|-1 \choose r-1}$, is known by all nodes whose indices are in $\mathcal{S}_{(k)}[i]$, and needed by all nodes whose indices are in ${\cal S} \backslash \mathcal{S}_{(k)}[i]$.

\subsubsection{Encoding}
The shuffling scheme of CDC consists of multiple rounds, each corresponding to all subsets of the $K$ nodes with a particular size. Within each subset, each node multicasts \emph{linear combinations} of the segments that are associated with it to the other nodes in the subset. More specifically, for each subset $\mathcal{S} \subseteq \{1,\ldots,K\}$ of size $\max\{r+1,s\} \leq |\mathcal{S}| \leq \min\{r+s,K\}$, we define $n_1 \triangleq {|{\cal S}|-1 \choose r-1}$ and $n_2 \triangleq {|{\cal S}|-2 \choose r-1}$. Then for each $k \in {\cal S}$, Node $k$ computes $n_2$ message symbols, denoted by $X_{k}^{\cal S}[1], X_{k}^{\cal S}[2],\ldots,X_{k}^{\cal S}[n_2]$ as follows. For some coefficients $\alpha_1,\ldots,\alpha_{n_1}$ where $\alpha_i \in \mathbb{F}_{2^{{r\choose |{\cal S}| -s}\frac{\eta_1\eta_2T}{r}}}$ for all $i=1,\ldots,n_1$, Node~$k$ computes
\begin{equation}
\begin{aligned}
X_{k}^{\cal S}[1] &\!=\! U_{\mathcal{S}_{(k)}[1],k}^{{\cal S} \backslash \mathcal{S}_{(k)}[1]} + U_{\mathcal{S}_{(k)}[2],k}^{{\cal S} \backslash \mathcal{S}_{(k)}[2]} + \cdots + U_{\mathcal{S}_{(k)}[n_1],k}^{{\cal S} \backslash \mathcal{S}_{(k)}[n_1]},\\
X_{k}^{\cal S}[2] &\!=\! \alpha_1 U_{\mathcal{S}_{(k)}[1],k}^{{\cal S} \backslash \mathcal{S}_{(k)}[1]} \!+\! \alpha_2 U_{\mathcal{S}_{(k)}[2],k}^{{\cal S} \backslash \mathcal{S}_{(k)}[2]} \!+ \cdots \!+\! \alpha_{n_1}\! U_{\mathcal{S}_{(k)}[n_1],k}^{{\cal S} \backslash \mathcal{S}_{(k)}[n_1]},\\
&\vdots\\
X_{k}^{\cal S}[n_2] &\!=\! \alpha_1^{n_2-1} U_{\mathcal{S}_{(k)}[1],k}^{{\cal S} \backslash \mathcal{S}_{(k)}[1]} + \alpha_2^{n_2-1} U_{\mathcal{S}_{(k)}[2],k}^{{\cal S} \backslash \mathcal{S}_{(k)}[2]} \\
&+ \cdots + \alpha_{n_1}^{n_2-1} U_{\mathcal{S}_{(k)}[n_1],k}^{{\cal S} \backslash \mathcal{S}_{(k)}[n_1]},
\end{aligned}
\end{equation}
or equivalently,
\begin{align}\label{eq:encoding}
\begin{bmatrix}
  X_{k}^{\cal S}[1]\\
  X_{k}^{\cal S}[2]\\
  \vdots\\
 X_{k}^{\cal S}[n_2] 
\end{bmatrix}
\!\!=\!\!
\underbrace{\begin{bmatrix}
  1 & 1 & \cdots & 1 \\
  \alpha_1 & \alpha_2 & \cdots & \alpha_{n_1} \\
  \vdots  & \vdots  & \ddots & \vdots  \\
  \alpha_1^{n_2-1} & \alpha_2^{n_2-1} & \cdots & \alpha_{n_1}^{n_2-1}
 \end{bmatrix}}_{{\bf A}^{\cal S}}
 \begin{bmatrix}
   U_{\mathcal{S}_{(k)}[1],k}^{{\cal S} \backslash \mathcal{S}_{(k)}[1]}\\
   U_{\mathcal{S}_{(k)}[2],k}^{{\cal S} \backslash \mathcal{S}_{(k)}[2]}\\
   \vdots \\
  U_{\mathcal{S}_{(k)}[n_1],k}^{{\cal S} \backslash \mathcal{S}_{(k)}[n_1]} 
 \end{bmatrix}.
\end{align}

We note that the above encoding process is the same at all nodes whose indices are in ${\cal S}$, i.e., each of them multiplies the same matrix ${\bf A}^{\cal S}$ in (\ref{eq:encoding}) with the segments associated with it.

Having generated the above message symbols, Node~$k$ multicasts them to the other nodes whose indices are in ${\cal S}$. 

\begin{remark}
When $s=1$, i.e., every output function is computed by one node, the above shuffling scheme only takes one round for all subsets ${\cal S}$ of size $|{\cal S}|=r+1$. Instead of multicasting linear combinations, every node in ${\cal S}$ can simply multicast the bit-wise XOR of its associated segments to the other $r$ nodes in ${\cal S}$. $\hfill \square$
\end{remark}

\subsubsection{Decoding}
For $j \in {\cal S}$ and $j \neq k$, there are a total of ${|{\cal S}|-2 \choose r-2}$ subsets of ${\cal S}$ that have size $r$ and simultaneously contain $j$ and $k$. Hence, among all $n_1$ segments  $U_{\mathcal{S}_{(k)}[1],k}^{{\cal S} \backslash \mathcal{S}_{(k)}[1]}, U_{\mathcal{S}_{(k)}[2],k}^{{\cal S} \backslash \mathcal{S}_{(k)}[2]},\ldots, U_{\mathcal{S}_{(k)}[n_1],k}^{{\cal S} \backslash \mathcal{S}_{(k)}[n_1]}$ associated with Node~$k$, ${|{\cal S}|-2 \choose r-2}$ of them are already known at Node~$j$, and the rest of $n_1 - {|{\cal S}|-2 \choose r-2}={|{\cal S}|-1 \choose r-1}-{|{\cal S}|-2 \choose r-2} = {|{\cal S}|-2 \choose r-1}=n_2$ segments are needed by Node~$j$. We denote the indices of the subsets that contain the element $k$ but not the element $j$ as $b_{jk}^1, b_{jk}^2,\ldots,b_{jk}^{n_2}$, such that $1 \leq b_{jk}^1 < b_{jk}^2<\cdots<b_{jk}^{n_2} \leq n_1$, and $j \notin {\cal S}_{(k)}[b_{jk}^i]$ for all $i=1,2,\ldots,n_2$. 

After receiving the symbols $X_{k}^{\cal S}[1], X_{k}^{\cal S}[2],\ldots,X_{k}^{\cal S}[n_2]$ from Node~$k$, Node~$j$ first removes the locally known segments from the linear combinations to generate $n_2$ symbols $Y_{jk}^{\cal S}[1],Y_{jk}^{\cal S}[2],\ldots, Y_{jk}^{\cal S}[n_2]$, such that
\begin{align}\label{eq:decoding}
\begin{bmatrix}
  Y_{jk}^{\cal S}[1]\\
  Y_{jk}^{\cal S}[2]\\
  \vdots\\
 Y_{jk}^{\cal S}[n_2] 
\end{bmatrix}
\!\!\!=\!\!\!
\underbrace{\begin{bmatrix}
  1 & 1 & \cdots & 1 \\
  \alpha_{b^1_{jk}} & \alpha_{b^2_{jk}} & \cdots & \alpha_{b^{n_2}_{jk}} \\
  \vdots  & \vdots  & \ddots & \vdots  \\
  \alpha^{n_2-1}_{b^1_{jk}} & \alpha^{n_2-1}_{b^2_{jk}} & \cdots & \alpha^{n_2-1}_{b^{n_2}_{jk}}
 \end{bmatrix}}_{{\bf B}_{jk}^{\cal S}}\!
 \begin{bmatrix}
   U_{\mathcal{S}_{(k)}[b_{jk}^1],k}^{{\cal S} \backslash \mathcal{S}_{(k)}[b_{jk}^1]}\\
   U_{\mathcal{S}_{(k)}[b_{jk}^2],k}^{{\cal S} \backslash \mathcal{S}_{(k)}[b_{jk}^2]}\\
   \vdots \\
  U_{\mathcal{S}_{(k)}[b_{jk}^{n_2}],k}^{{\cal S} \backslash \mathcal{S}_{(k)}[b_{jk}^{n_2}]} 
 \end{bmatrix}\!,
\end{align}
where ${\bf B}_{jk}^{\cal S} \in \mathbb{F}_{2^{{r\choose |{\cal S}| -s}\frac{\eta_1\eta_2T}{r}}}^{n_2 \times n_2}$ is a square sub-matrix of ${\bf A}^{\cal S}$ in (\ref{eq:encoding}) that contains the columns with indices $b_{jk}^1, b_{jk}^2,\ldots,b_{jk}^{n_2}$ of ${\bf A}_k^{\cal S}$.  

Node $j$ can decode the desired segments from Node $k$ if the matrix ${\bf B}_{jk}^{\cal S}$ is invertible. We note that ${\bf B}_{jk}^{\cal S}$ is a Vandermonde matrix, and it is invertible if $\alpha_{b^1_{jk}}, \alpha_{b^2_{jk}}, \ldots,\alpha_{b^{n_2}_{jk}}$ are all distinct. This holds for all $j \in {\cal S} \backslash \{k\}$ if there exist $n_1$ distinct coefficients in $\mathbb{F}_{2^{{r\choose |{\cal S}| -s}\frac{\eta_1\eta_2T}{r}}}$, which requires $2^{{r\choose |{\cal S}| -s}\frac{\eta_1\eta_2T}{r}} \geq n_1 = {|{\cal S}|-1 \choose r-1}$, or equivalently $T \geq \frac{r\log {|{\cal S}|-1 \choose r-1}}{{r \choose |{\cal S}|-s}\eta_1 \eta_2}$. Finally, the proposed coded shuffling scheme can successfully deliver all the required intermediate values within all subsets ${\cal S}$ with $\max\{r+1,s\} \leq |\mathcal{S}| \leq \min\{r+s,K\}$, if $T$ is sufficiently large, i.e.,  
\begin{align}
T \geq \max_{\max\{r+1,s\} \leq |\mathcal{S}| \leq \min\{r+s,K\}} \frac{r\log {|{\cal S}|-1 \choose r-1}}{{r \choose |{\cal S}|-s}\eta_1 \eta_2}.
\end{align}

\subsection{Correctness of CDC}
We demonstrate the correctness of the above shuffling scheme by showing that after the Shuffle phase, each node can decode all of the required intermediate values to compute its assigned Reduce functions. We use Node 1 as an example, and similar arguments apply to all other nodes. WLOG we assume that the Reduce function $h_1$ is to be computed by Node 1. Node 1 will need a total of ${K-1 \choose r}\eta_1$ distinct intermediate values of $h_1$ from other nodes (it already knows $\frac{r\bar{N}}{K} = \bar{N} - {K-1 \choose r}\eta_1$ intermediate values of $h_1$ by mapping the files in ${\cal M}_1$). By the assignment of the Reduce functions, there exits a subset $\mathcal{S}_2$ of size $s$ containing Node 1 such that all nodes in $\mathcal{S}_2$ need to compute $h_1$. Then, during the data shuffling process within each subset $\mathcal{S}$ containing $\mathcal{S}_2$ (note that by the definition of ${\cal V}_{{\cal S}_1}^{{\cal S} \backslash {\cal S}_1}$ in (\ref{eq:V}), the intermediate values of $h_1$ will \emph{not} be communicated to Node 1 if ${\cal S}_2 \nsubseteq {\cal S}$, and this is because that some node outside ${\cal S}$ also wants to compute $h_1$), there are ${s-1 \choose |{\cal S}|-r-1}$ subsets ${\cal S}_1$ of ${\cal S}$ with size $|{\cal S}_1|=r$ such that $1 \notin {\cal S}_1$ and ${\cal S} \backslash {\cal S}_1 \subseteq {\cal S}_2$, and thus Node 1 decodes ${s-1 \choose |\mathcal{S}|-r-1}\eta_1$ distinct intermediate values of $h_1$. Therefore, the total number of distinct intermediate values of $h_1$ Node 1 decodes over the entire Shuffle phase is
\begin{align}
\sum\limits_{\ell = \max\{r+1,s\}}^{\min\{r+s,K\}} \!\!  {s-1 \choose \ell -r-1}  \!\! {K - s \choose \ell-s}\eta_1 \!=\! {K-1 \choose r}\eta_1,
\end{align}
which matches the required number of intermediate values for $h_1$. This is also true for all the other Reduce functions assigned to Node 1. 

\subsection{Communication Load}
In the above shuffling scheme, for each subset $\mathcal{S} \subseteq \{1,\ldots,K\}$ of size $\max\{r+1,s\} \leq |\mathcal{S}| \leq \min\{r+s,K\}$, each Node $k \in {\cal S}$ communicates $n_2 = {|{\cal S}|-2 \choose r-1}$ message symbols. Each of these symbols contains ${r\choose |{\cal S}| -s}\frac{\eta_1\eta_2 T}{r}$ bits. Hence, all nodes whose indices are in ${\cal S}$ communicate a total of $|{\cal S}|{|{\cal S}| -2 \choose r -1}{r\choose |{\cal S}| -s}\frac{\eta_1\eta_2 T}{r}$ bits. The overall communication load achieved by the proposed CDC scheme is 
\begin{align}
L_{\textup{coded}}(r,s) &= \lim_{N \rightarrow \infty}\sum \limits_{\ell = \max\{r+1,s\}}^{\min \{r+s,K\}} \!\! \frac{{K\choose \ell} \! \frac{\ell}{r}\! {\ell -2 \choose r -1}\! {r\choose \ell -s }\eta_1\eta_2T}{QNT} \nonumber \\
&= \lim_{N \rightarrow \infty} \sum \limits_{\ell = \max\{r+1,s\}}^{\min \{r+s,K\}} \!\! \frac{\ell {K \choose \ell} {\ell -2 \choose r -1} {r \choose \ell -s}\bar{N}}{r {K \choose r}{K \choose s}N} \nonumber\\
&=\sum \limits_{\ell = \max\{r+1,s\}}^{\min \{r+s,K\}} \!\! \frac{\ell {K \choose \ell} {\ell -2 \choose r -1} {r \choose \ell -s}}{r {K \choose r}{K \choose s}}.\label{eq:s}
\end{align}

\subsection{Non-Integer Valued Computation Load}
For non-integer valued computation load $r \geq 1$, we generalize the CDC scheme as follows. We first expand the computation load $r = \alpha r_1 + (1-\alpha)r_2$ as a convex combination of $r_1 \triangleq \lfloor r \rfloor$ and $r_2 \triangleq \lceil r \rceil$, for some $0 \leq \alpha \leq 1$. Then we partition the set of $\bar{N}$ input files $\{w_1,\ldots,w_{\bar{N}}\}$ into two disjoint subsets $\mathcal{I}_1$ and $\mathcal{I}_2$ of sizes $|\mathcal{I}_1| = \alpha \bar{N}$ and $|\mathcal{I}_2| = (1-\alpha) \bar{N}$. We next apply the CDC scheme described above respectively to the files in $\mathcal{I}_1$ with a computation load $r_1$ and the files in $\mathcal{I}_2$ with a computation load $r_2$, to compute each of the $Q$ output functions at the same set of $s$ nodes. This results in a communication load of 
\begin{align}
&\lim_{N \rightarrow \infty}\frac{Q \alpha \bar{N} L_{\textup{coded}}(r_1,s) T + Q (1-\alpha) \bar{N} L_{\textup{coded}}(r_2,s)T}{QNT} \nonumber \\
=& \alpha L_{\textup{coded}}(r_1,s) + (1-\alpha) L_{\textup{coded}}(r_2,s),
\end{align} 
where $L_{\textup{coded}}(r,s)$ is the communication load achieved by CDC in (\ref{eq:s}) for integer-valued $r,s \in \{1,\ldots,K\}$.

Using this generalized CDC scheme, for any two integer-valued computation loads $r_1$ and $r_2$, the points on the line segment connecting $(r_1,L_{\textup{coded}}(r_1,s))$ and $(r_2,L_{\textup{coded}}(r_2,s))$ are achievable. Therefore, for general $1\leq r \leq K$, the \emph{lower convex envelop} of the achievable points $\{(r,L_{\textup{coded}}(r,s)): r\in \{1,\ldots,K\}\}$ is achievable. This proves the upper bound on the computation-communication function in Theorem~2 (also the achievability part of Theorem~1 by setting $s=1$).

\begin{remark}
The ideas of efficiently creating and exploiting coded multicasting opportunities have been introduced in caching problems~\cite{maddah2014fundamental,maddah2013decentralized,ji2014fundamental}. In this section, we illustrated how coding opportunities can be utilized in distributed computing to slash the load of communicating intermediate values, by designing a particular assignment of extra computations across distributed computing nodes. We note that the calculated intermediate values in the Map phase mimics the locally stored cache contents in caching problems, providing the ``side information'' to enable coding in the following Shuffle phase (or content delivery). 

For the case of $s=1$ where no two nodes are interested in computing a common Reduce function, the coded data shuffling of CDC is similar to a coded transmission strategy in wireless D2D networks proposed in~\cite{ji2014fundamental}, where the side information enabling coded multicasting are pre-fetched in a specific repetitive manner in the caches of wireless nodes (in CDC such information is obtained by computing the Map functions locally). When $s$ is larger than $1$, i.e., every Reduce function needs to be computed at multiple nodes, our CDC scheme creates novel coding opportunities that exploit both the redundancy of the Map computations and the commonality of the data requests for Reduce functions across nodes, further reducing the communication load. $\hfill \square$
\end{remark}

\begin{remark}
Generally speaking, we can view the Shuffle phase of the considered distributed computing framework as an instance of the index coding problem~\cite{birk2006coding,bar2011index}, in which a central server aims to design a broadcast message (code) with minimum length to simultaneously satisfy the requests of all the clients, given the clients' side information stored in their local caches. Note that while a randomized linear network coding approach (see e.g., \cite{ahlswede2000network,KM03,HKMKE03}) is sufficient to implement any multicast communication where messages are intended by all receivers, it is generally sub-optimal for index coding problems where every client requests different messages. Although the index coding problem is still open in general, for the considered distributed computing scenario where we are given the flexibility of designing Map computation (thus the flexibility of designing side information), we prove in the next  two sections \emph{tight} lower bounds on the minimum communication loads for the cases $s=1$ and $s > 1$ respectively, demonstrating the optimality of the proposed CDC scheme.
$\hfill \square$
\end{remark}

\section{Converse of Theorem 1}
In this section, we prove the lower bound on $L^*(r)$ in Theorem~1. 

For $k \in \{1,\ldots,K\}$, we denote the set of indices of the files mapped by Node~$k$ as $\mathcal{M}_k$, and the set of indices of the Reduce functions computed by Node~$k$ as $\mathcal{W}_k$. As the first step, we consider the communication load for a given file assignment $\mathcal{M} \triangleq ({\cal M}_1,{\cal M}_2\ldots,{\cal M}_K)$ in the Map phase. We denote the minimum communication load under the file assignment $\mathcal{M}$ by $L^*_\mathcal{M}$.

We denote the number of files that are mapped at $j$ nodes under a file assignment ${\cal M}$, as $a^j_{{\cal M}}$, for all $j \in \{1,\ldots,K\}$:
\begin{equation} \label{eq:count}
a^j_{{\cal M}} = \sum \limits_{{\cal J} \subseteq \{1,\ldots,K\}: |{\cal J}|=j} |(\underset{k \in {\cal J}}{\cap} {\cal M}_k) \backslash (\underset{i \notin {\cal J}}{\cup} {\cal M}_i )|.
\end{equation}

\begin{figure}[htbp]
   \centering
   \includegraphics[width=0.48\textwidth]{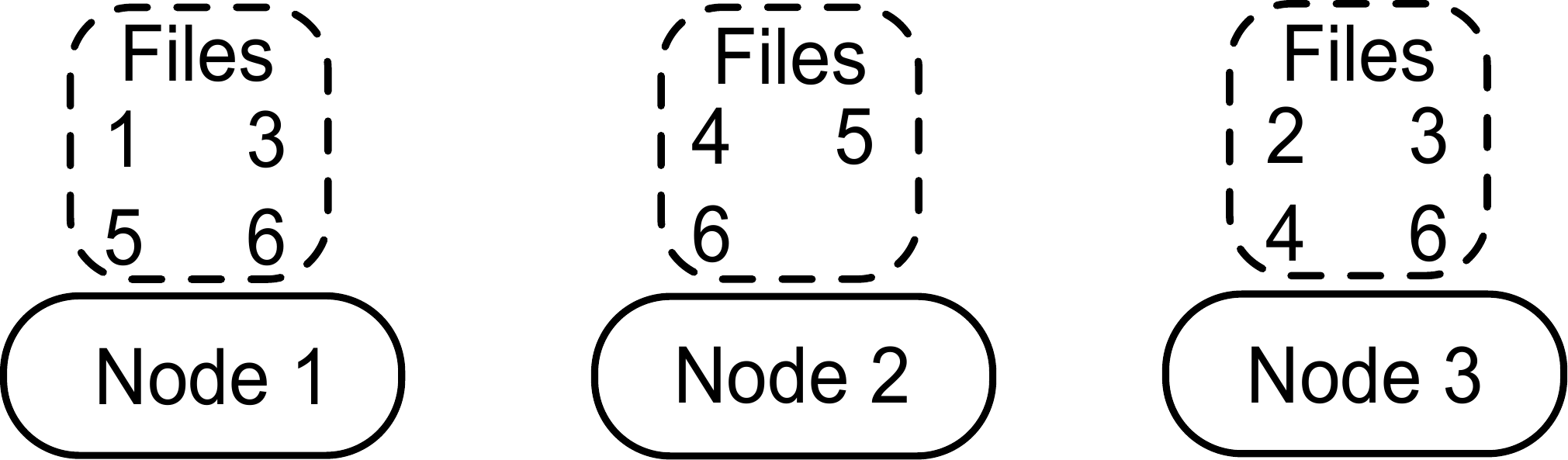}
   \caption{A file assignment for $N=6$ files and $K=3$ nodes.}
   \label{fig:ajm}
\end{figure}

For example, for the particular file assignment in Fig.~\ref{fig:ajm}, i.e., ${\cal M} = (\{1,3,5,6\}, \{4,5,6\}, \{2,3,4,6\})$, $a^1_{\cal M}=2$ since File 1 and File 2 are mapped on a single node (i.e., Node 1 and Node 3 respectively). Similarly, we have $a^2_{\cal M} = 3$ (Files 3, 4, and 5), and $a^3_{\cal M} = 1$ (File 6).

For a particular file assignment ${\cal M}$, we present a lower bound on $L^*_\mathcal{M}$ in the following lemma.
\begin{lemma}
	
$L^*_\mathcal{M} \geq \sum\limits_{j=1}^{K} \frac{a^{j}_{\cal M}}{N}\cdot\frac{K-j}{Kj}$.

\end{lemma} 

Next, we first demonstrate the converse of Theorem~1 using Lemma~1, and then give the proof of Lemma~1. 

\noindent \emph{Converse Proof of Theorem~1.} It is clear that the minimum communication load $L^*(r)$ is lower bounded by the minimum value of $L^*_\mathcal{M}$ over all possible file assignments which admit a computation load of $r$:
\begin{equation}
L^*(r) \geq \underset{{\cal M}: |{\cal M}_1| + \cdots+|{\cal M}_K|=rN}{\inf} L^*_\mathcal{M}.
\end{equation}

Then by Lemma~1, we have 
\begin{align}\label{eq:lower1}
L^*(r)\geq \inf_{{\cal M}: |{\cal M}_1| + \cdots+|{\cal M}_K|=rN}\sum_{j=1}^{K} \frac{a^{j}_{\cal M}}{N} \cdot \frac{K-j}{Kj}.
\end{align}

For every file assignment ${\cal M}$ such that $|{\cal M}_1| + \cdots+|{\cal M}_K|=rN$, $\{a^j_{\cal M}\}_{j=1}^K$ satisfy
\begin{align}
a^j_{\cal M} &\geq 0, \ j\in\{1,...,K\}, \label{eq:positive}\\
\sum_{j=1}^{K} a^j_{\cal M}&=N, \label{eq:convex} \\ 
\sum_{j=1}^{K} j a^j_{\cal M} &=rN. \label{eq:compute}
\end{align}

Then since the function $\frac{K-j}{Kj}$ in (\ref{eq:lower1}) is convex in $j$, and by (\ref{eq:convex}) $\sum\limits_{j=1}^{K} \frac{a^j_{\cal M}}{N} =1$, (\ref{eq:lower1}) becomes
\begin{align}\label{eq:general}
L^*(r)\geq \inf_{{\cal M}: |{\cal M}_1| + \cdots+|{\cal M}_K|=rN}  \frac{K-\sum\limits_{j=1}^K j\frac{a^j_{\cal M}}{N}}{K\sum\limits_{j=1}^K j\frac{a^j_{\cal M}}{N}} \overset{(a)}{=}\frac{K-r}{Kr} , 
\end{align}
where (a) is due to the requirement imposed by the computation load in (\ref{eq:compute}). 

The lower bound on $L^*(r)$ in (\ref{eq:general}) holds for general $1 \leq r \leq K$. We can further improve the lower bound for non-integer valued $r$ as follows. For a particular $r \notin \mathbb{N}$, we first find the line $p+qj$ as a function of $1 \leq j \leq K$ connecting the two points $(\lfloor r \rfloor, \frac{K-\lfloor r \rfloor}{K\lfloor r \rfloor})$ and $(\lceil r \rceil, \frac{K-\lceil r \rceil}{K\lceil r \rceil})$. More specifically, we find $p,q \in \mathbb{R}$ such that 
\begin{align}
p+q j|_{j=\lfloor r \rfloor} =  \frac{K-\lfloor r \rfloor}{K\lfloor r \rfloor},\\
p+q j|_{j=\lceil r \rceil} =  \frac{K-\lceil r \rceil}{K\lceil r \rceil}.
\end{align}

Then by the convexity of the function $\frac{K-j}{Kj}$ in $j$, we have for integer-valued $j=1,\ldots,K$,
\begin{equation}
\frac{K-j}{Kj} \geq p + qj, \quad j=1,\ldots,K.
\end{equation}

Then (\ref{eq:lower1}) reduces to 
\begin{align}
L^*(r)&\geq \inf_{{\cal M}: |{\cal M}_1| + \cdots+|{\cal M}_K|=rN}\sum_{j=1}^{K} \frac{a^{j}_{\cal M}}{N} \cdot (p+qj)\\
&= \inf_{{\cal M}: |{\cal M}_1| + \cdots+|{\cal M}_K|=rN}   \sum_{j=1}^{K} \frac{a^{j}_{\cal M}}{N} \cdot p +  \sum_{j=1}^{K} \frac{ja^{j}_{\cal M}}{N} \cdot q\\
& \overset{(b)}{=} p+qr,
\end{align}
where (b) is due to the constraints on $\{a^j_{\cal M}\}_{j=1}^K$ in (\ref{eq:convex}) and (\ref{eq:compute}).

Therefore, $L^*(r)$ is lower bounded by the lower convex envelop of the points $\{(r,\frac{K-r}{Kr}):r\in\{1,...,K\}\}$. This completes the proof of the converse part of Theorem~1.
 $\hfill \blacksquare$
 
\begin{remark}
	Although the model proposed in this paper only allows each node sending messages independently, we can show that even if the data shuffling process can be carried out in multiple rounds and dependency between messages are allowed, the lower bound on $L^*(r)$ remains the same. $\hfill \square$
\end{remark}

We devote the rest of this section to the proof of Lemma~1. To prove Lemma~1, we develop a lower bound on the number of bits communicated by any subset of nodes, by induction on the size of the subset. In particular, for a subset of computing nodes, we first characterize a lower bound on the minimum number of bits required by a particular node in the subset, which is given by a cut-set bound separating this node and all the other nodes in the subset. Then, we combine this bound with the lower bound on the number of bits communicated by the rest of the nodes in the subset, which is given by the inductive argument. 

\noindent \emph{Proof of Lemma~1.} For $q\in\{1,...,Q\}$, $n\in\{1,...,N\}$, we let $V_{q,n}$ be i.i.d. random variables uniformly distributed on $\mathbb{F}_{2^T}$. We let the intermediate values $v_{q,n}$ be the realizations of $V_{q,n}$. For some ${\cal Q} \subseteq \{1,\ldots,Q\}$ and ${\cal N} \subseteq \{1,\ldots,N\}$, we define 
\begin{equation}
V_{{\cal Q}, \,{\cal N}} \triangleq \{V_{q,n}: q \in {\cal Q}, n \in {\cal N}\}.
\end{equation}

Since each message $X_k$ is generated as a function of the intermediate values that are computed at Node $k$, the following equation holds for all $k\in\{1,...,K\}$.
 	\begin{align}
 	H(X_k|V_{:,{\mathcal{M}_{k}}})=0, \label{causality}
 	\end{align}
where we use ``:'' to denote the set of all possible indices.

The validity of the shuffling scheme requires that for all $k\in\{1,...,K\}$, the following equation holds :
\begin{align}
H(V_{\mathcal{W}_k,:}|X_:,V_{:,{\mathcal{M}_k}})=0.
\label{validity}
\end{align}

For a subset $\mathcal{S}\subseteq \{1,...,K\}$, we define 
\begin{equation}\label{eq:rest}
Y_{\mathcal{S}} \triangleq (V_{\mathcal{W}_{\mathcal{S}},:},V_{:,\mathcal{M}_{\mathcal{S}}}),
 \end{equation}
 which contains all the intermediate values required by the nodes in ${\cal S}$ and all the intermediate values known locally by the nodes in ${\cal S}$ after the Map phase.
 
For any subset ${\cal S} \subseteq \{1,\ldots,K\}$ and a file assignment ${\cal M}$, we denote the number of files that are \emph{exclusively} mapped by $j$ nodes in $\mathcal{S}$ as $a^{j,\mathcal{S}}_{\cal M}$:
\begin{equation}\label{eq:countSub}
a^{j,{\cal S}}_{{\cal M}} \triangleq \sum \limits_{{\cal J} \subseteq {\cal S}: |{\cal J}|=j} |(\underset{k \in {\cal J}}{\cap} {\cal M}_k) \backslash (\underset{i \notin {\cal J}}{\cup} {\cal M}_i )|,
\end{equation}
and the message symbols communicated by the nodes whose indices are in ${\cal S}$ as 
\begin{equation}
X_{\cal S} = \{X_k: k \in {\cal S}\}.
\end{equation}

Then we prove the following claim.
\begin{claim}
\noindent For any subset $\mathcal{S}\subseteq \{1,...,K\}$, we have
\begin{equation}
H(X_{\mathcal{S}}|Y_{\mathcal{S}^c})\geq T\sum\limits_{j=1}^{|\mathcal{S}|} a^{j,\mathcal{S}}_{\cal M}\frac{Q}{K}\cdot\frac{|\mathcal{S}|-j}{j},
\end{equation}
where ${\cal S}^c \triangleq \{1,\ldots,K\} \backslash {\cal S}$ denotes the complement of ${\cal S}$. $\hfill \square$
\end{claim}

We prove Claim~1 by induction.

a. If $\mathcal{S} = \{k\}$ for any $k \in \{1,\ldots,K\}$, obviously
\begin{align}
H(X_k|Y_{\{1,\ldots,K\}\backslash \{k\}})&\geq 0	=T a^{1,\{k\}}_{\cal M}\frac{Q}{K}\cdot\frac{1-1}{1}.
\end{align}

b. Suppose the statement is true for all subsets of size $S_0$. 

For any $\mathcal{S}\subseteq \{1,...,K\}$ of size $|\mathcal{S}|=S_0+1$ and any $k \in \mathcal{S}$, we have
 	\begin{align}
 		&H(X_{\mathcal{S}}|Y_{\mathcal{S}^c}) \nonumber\\
 		=&\frac{1}{|\mathcal{S}|} \sum_{k\in\mathcal{S}}H(X_{\mathcal{S}},X_k|Y_{\mathcal{S}^c})\\
 		=&\frac{1}{|\mathcal{S}|} \sum_{k\in\mathcal{S}} (H(X_{\mathcal{S}}|X_k,  Y_{\mathcal{S}^c})+H(X_k| Y_{\mathcal{S}^c}))\\
 		\geq&\frac{1}{|\mathcal{S}|} \sum_{k\in\mathcal{S}} H(X_{\mathcal{S}}|X_k,  Y_{\mathcal{S}^c})+\frac{1}{|\mathcal{S}|}H(X_\mathcal{S}| Y_{\mathcal{S}^c}).\label{eq:XS}
 	\end{align}

From (\ref{eq:XS}), we have
 	\begin{align}
 	H(X_{\mathcal{S}}|Y_{\mathcal{S}^c})&\geq\frac{1}{|\mathcal{S}|-1} \sum_{k\in\mathcal{S}} H(X_{\mathcal{S}}|X_k, Y_{\mathcal{S}^c})\\
 	&\geq\frac{1}{S_0} \sum_{k\in\mathcal{S}} H(X_{\mathcal{S}}|X_k, V_{:,{\mathcal{M}_k}}, Y_{\mathcal{S}^c})\\
 	&=\frac{1}{S_0} \sum_{k\in\mathcal{S}} H(X_{\mathcal{S}}| V_{:,{\mathcal{M}_k}}, Y_{\mathcal{S}^c}). \label{eq:all}
 	\end{align}
	
For each $k \in {\cal S}$, we have the following subset version of (\ref{causality}) and (\ref{validity}).
 	\begin{align}
 	H(X_k|V_{:,{\mathcal{M}_{k}}},Y_{\mathcal{S}^c})&=0, \label{scausality}\\
 	H(V_{\mathcal{W}_k,:}|X_\mathcal{S},V_{:,{\mathcal{M}_k}},Y_{\mathcal{S}^c})&=0. \label{svalidity}
 	\end{align}
 	
 	Consequently, 
 	\begin{align}
 	&H(X_\mathcal{S}, V_{\mathcal{W}_k,:}|V_{:,{\mathcal{M}_k}},Y_{\mathcal{S}^c}) = H(X_\mathcal{S}|V_{:,{\mathcal{M}_k}},Y_{\mathcal{S}^c})\\
	=&H(V_{\mathcal{W}_k,:}|V_{:,{\mathcal{M}_k}},Y_{\mathcal{S}^c})+H(X_\mathcal{S}|V_{\mathcal{W}_k,:},V_{:,{\mathcal{M}_k}},Y_{\mathcal{S}^c}).
 	\label{intermediate}
 	\end{align}
	
The first term on the RHS of (\ref{intermediate}) can be lower bounded as follows.
\begin{align}
H(V_{\mathcal{W}_k,:}|V_{:,{\mathcal{M}_k}},Y_{\mathcal{S}^c})&= H(V_{\mathcal{W}_k,:}|V_{:,{\mathcal{M}_k}},V_{\mathcal{W}_{\mathcal{S}^c},:},V_{:,\mathcal{M}_{\mathcal{S}^c}})\nonumber\\
&\overset{(a)}{=}H(V_{\mathcal{W}_k,:}|V_{:,{\mathcal{M}_k}},V_{:,\mathcal{M}_{\mathcal{S}^c}})\\
&\overset{(b)}{=}H(V_{\mathcal{W}_k,:}|V_{{\cal W}_k,{\mathcal{M}_k}\cup{\cal M}_{{\cal S}^c}})\\
&\overset{(c)}{=}\!\!\sum_{q \in {\cal W }_k} H(V_{\{q\},:}|V_{\{q\},{\mathcal{M}_k}\cup{\cal M}_{{\cal S}^c}})\\
&\overset{(d)}{=}\frac{Q}{K}T\sum_{j=0}^{S_0}  a^{j,\mathcal{S}\backslash \{k\}}_{\cal M}\\
&\geq \frac{Q}{K}T\sum_{j=1}^{S_0}  a^{j,\mathcal{S}\backslash \{k\}}_{\cal M}, \label{eq:first}
\end{align} 
where (a) is due to the independence of intermediate values and the fact that ${\cal W}_k \cap {\cal W}_{{\cal S}^c} = \emptyset$ (different nodes calculate different output functions), (b) and (c) are due to the independence of intermediate values, and (d) is due to the independence of the intermediate values and the fact that $|{\cal W}_k|=\frac{Q}{K}$.	
 	
The second term on the RHS of (\ref{intermediate}) can be lower bounded by the induction assumption:
    \begin{align}
    H(X_\mathcal{S}|V_{\mathcal{W}_k,:},V_{:,{\mathcal{M}_k}},Y_{\mathcal{S}^c})&=H(X_{\mathcal{S}\backslash \{k\}}|Y_{(\mathcal{S}\backslash \{k\})^c})\\    
    &\geq T\sum_{j=1}^{S_0} a^{j,\mathcal{S}\backslash \{k\}}_{\cal M}\frac{Q}{K}\cdot\frac{S_0-j}{j}. \label{eq:second}
    \end{align} 

Thus by (\ref{eq:all}), (\ref{intermediate}), (\ref{eq:first}) and (\ref{eq:second}), we have
	\begin{align}
	H(X_{\mathcal{S}}|Y_{\mathcal{S}^c})\geq&\frac{1}{S_0} \sum_{k\in\mathcal{S}} H(X_{\mathcal{S}}| V_{:,{\mathcal{M}_k}}, Y_{\mathcal{S}^c})\\
	=& \frac{1}{S_0} \sum_{k\in\mathcal{S}} \bigg(H(V_{\mathcal{W}_k,:}|V_{:,{\mathcal{M}_k}},Y_{\mathcal{S}^c}) \nonumber\\
	&+H(X_\mathcal{S}|V_{\mathcal{W}_k,:},V_{:,{\mathcal{M}_k}},Y_{\mathcal{S}^c})\bigg) \label{eq:twoParts}\\
	\geq& \frac{1}{S_0} \sum_{k\in\mathcal{S}} \bigg(T\sum_{j=1}^{S_0} a^{j,\mathcal{S}\backslash \{k\}}_{\cal M} \frac{Q}{K} \nonumber\\
	&+ T\sum_{j=1}^{S_0} a^{j,\mathcal{S}\backslash \{k\}}_{\cal M}\frac{Q}{K}\cdot\frac{S_0-j}{j}\bigg)\\ 
	=& \frac{T}{S_0} \sum_{k\in\mathcal{S}} \sum_{j=1}^{S_0} a^{j,\mathcal{S}\backslash \{k\}}_{\cal M}\frac{Q}{K}\cdot\frac{S_0}{j}\\
	=& T \sum_{j=1}^{S_0} \frac{Q}{K}\cdot \frac{1}{j} \sum_{k\in\mathcal{S}}  a^{j,\mathcal{S}\backslash \{k\}}_{\cal M}. \label{eq:int1}
	\end{align}

By the definition of $a^{j,\mathcal{S}}_{\cal M}$, we have the following equations.
\begin{align}
&\sum_{k\in\mathcal{S}} \nonumber a^{j,\mathcal{S}\backslash \{k\}}_{\cal M}\\
=&\!\sum_{k\in\mathcal{S}} \sum_{n=1}^{N} \mathbbm{1}{(\textup{file $n$ is only mapped by some nodes in $\mathcal{S}\backslash \{k\}$})} \nonumber\\
& \times  \mathbbm{1}{(\textup{file $n$ is mapped by $j$ nodes})}\\
=&\sum_{n=1}^{N}  \mathbbm{1}{(\textup{file $n$ is only mapped by $j$ nodes in ${\cal S}$})} \nonumber \\
&\times \sum_{k\in\mathcal{S}}  \mathbbm{1}{(\textup{file $n$ is not mapped by Node $k$})}\\
=& \sum_{n=1}^{N}  \mathbbm{1}{(\textup{file $n$ is only mapped by $j$ nodes in ${\cal S}$})} (|\mathcal{S}|-j)\\
=& a^{j,\mathcal{S}}_{\cal M}(S_0+1-j). \label{eq:int2}
\end{align}

Applying (\ref{eq:int2}) to (\ref{eq:int1}) yields
\begin{align}
H(X_{\mathcal{S}}|Y_{\mathcal{S}^c})&\geq T\sum_{j=1}^{S_0}  a^{j,\mathcal{S}}_{\cal M} {\frac{Q}{K}} \cdot\frac{S_0+1-j}{j}\\
&=T\sum_{j=1}^{S_0+1}  a^{j,\mathcal{S}}_{\cal M} {\frac{Q}{K}} \cdot\frac{S_0+1-j}{j}.
\end{align}
	
c. Thus for all subsets $\mathcal{S}\subseteq \{1,...,K\}$, the following equation holds:
\begin{align}
H(X_{\mathcal{S}}|Y_{\mathcal{S}^c})\geq T\sum_{j=1}^{|\mathcal{S}|} a^{j,\mathcal{S}}_{\cal M}\frac{Q}{K}\cdot\frac{|\mathcal{S}|-j}{j},
\end{align}
which proves Claim~1.

Then by Claim~1, let $\mathcal{S}=\{1,...,K\}$ be the set of all $K$ nodes, 
\begin{align}
L^*_\mathcal{M}\geq \frac{H(X_{\mathcal{S}}|Y_{\mathcal{S}^c})}{QNT}\geq \sum_{j=1}^{K} \frac{a^{j}_{\cal M}}{N}\cdot\frac{K-j}{Kj}.
\end{align}

This completes the proof of Lemma~1. $\hfill \blacksquare$

\section{Converse of Theorem 2}
In this section, we prove the lower bound on $L^*(r,s)$ in Theorem~2, which generalizes the converse result of Theorem~1 for the case $s >1$. Since the lower bound on $L^*(r,1)$ in Theorem~2 exactly matches the lower bound on $L^*(r)$ in Theorem~1, we focus on the case $s >1$ (i.e., each Reduce function is calculated by 2 or more nodes) throughout this section.

We denote the minimum communication load under a particular file assignment $\mathcal{M}$ as $L^*_\mathcal{M}(s)$, and we present a lower bound on $L^*_\mathcal{M}(s)$ in the following lemma.
\begin{lemma}
$L^*_\mathcal{M}(s) \geq \sum\limits_{j=1}^{K} \frac{a^{j}_{\cal M}}{N}  \sum \limits_{\ell = \max\{j,s\}}^{\min\{j+s,K\}} \frac{{K-j \choose \ell-j}{j \choose \ell-s}}{{K \choose s}} \cdot \frac{\ell -j}{\ell-1}$, where $a^j_{\cal M}$ is defined in (\ref{eq:count}).
\end{lemma} 

In the rest of this section, we first prove the converse part of Theorem~2 by showing $L^*(r,s) \geq \sum \limits_{\ell = \max\{r,s\}}^{\min\{r+s,K\}} \frac{{K-r \choose \ell-r}{r \choose \ell-s}}{{K \choose s}} \cdot \frac{\ell -r}{\ell-1}$, and then give the proof of Lemma~2.

\noindent \emph{Converse Proof of Theorem~2.} The minimum communication load $L^*(r,s)$ is lower bounded by the minimum value of $L^*_\mathcal{M}(s)$ over all possible file assignments having a computation load of $r$:
\begin{equation}\label{eq:lower}
L^*(r,s) \geq \underset{{\cal M}: |{\cal M}_1| + \cdots+|{\cal M}_K|=rN}{\inf} L^*_\mathcal{M}(s).
\end{equation}

For every file assignment ${\cal M}$ such that $|{\cal M}_1| + \cdots+|{\cal M}_K|=rN$, $\{a^j_{\cal M}\}_{j=1}^K$ satisfy the same conditions as the case of $s=1$ in (\ref{eq:positive}), (\ref{eq:convex}) and (\ref{eq:compute}).

For a general computation load $1 \leq r \leq K$, and the function $L_{\textup{coded}}(r,s) = \sum \limits_{\ell = \max\{r+1,s\}}^{\min \{r+s,K\}} \!\! \frac{\ell {K \choose \ell} {\ell -2 \choose r -1} {r \choose \ell -s}}{r {K \choose r}{K \choose s}}= \sum \limits_{\ell = \max\{r,s\}}^{\min\{r+s,K\}} \frac{{K-r \choose \ell-r}{r \choose \ell-s}}{{K \choose s}} \cdot \frac{\ell -r}{\ell-1}$ as defined in (\ref{eq:s}), we first find the line $p+qj$ as a function of $1 \leq j \leq K$ connecting the two points $(\lfloor r \rfloor, L_{\textup{coded}}(\lfloor r \rfloor,s))$ and $(\lceil r \rceil, L_{\textup{coded}}(\lceil r \rceil,s))$. More specifically, we find $p,q \in \mathbb{R}$ such that 
\begin{align}
p+q j|_{j=\lfloor r \rfloor} =  L_{\textup{coded}}(\lfloor r \rfloor,s),\\
p+q j|_{j=\lceil r \rceil} = L_{\textup{coded}}(\lceil r \rceil,s).
\end{align}

Then by the convexity of the function $L_{\textup{coded}}(j,s)$ in $j$, we have for integer-valued $j = 1,\ldots,K$,
\begin{equation}\label{eq:convex-cascade}
L_{\textup{coded}}(j,s) = \sum \limits_{\ell = \max\{j,s\}}^{\min\{j+s,K\}} \frac{{K-j \choose \ell-j}{j \choose \ell-s}}{{K \choose s}} \cdot \frac{\ell -j}{\ell-1} \geq p + qj.
\end{equation}

Next, we first apply Lemma~2 to (\ref{eq:lower}), then by (\ref{eq:convex-cascade}), we have
\begin{align}
L^*(r,s)&\geq \inf_{{\cal M}: |{\cal M}_1| + \cdots+|{\cal M}_K|=rN}\sum_{j=1}^{K} \frac{a^{j}_{\cal M}}{N} \cdot (p+qj)\\
&= \inf_{{\cal M}: |{\cal M}_1| + \cdots+|{\cal M}_K|=rN}   \sum_{j=1}^{K} \frac{a^{j}_{\cal M}}{N} \cdot p +  \sum_{j=1}^{K} \frac{ja^{j}_{\cal M}}{N} \cdot q\\
& \overset{(a)}{=} p+qr,
\end{align}
where (a) is due to the constraints on $\{a^j_{\cal M}\}_{j=1}^K$ in (\ref{eq:convex}) and (\ref{eq:compute}).

Therefore, $L^*(r,s)$ is lower bounded by the lower convex envelop of the points $\{(r,L_{\textup{coded}}(r,s)):r\in\{1,...,K\}\}$. This completes the proof of the converse part of Theorem~2.
 $\hfill \blacksquare$
 
The proof of lemma~2 follows the same steps of the proof of Lemma~1, where a lower bound on the number of bits communicated by any subset of nodes, for the case of $s > 1$, is established by induction. 

\noindent \emph{Proof of Lemma~2.} We first prove the following claim.

\begin{claim}
\noindent For any subset $\mathcal{S}\subseteq \{1,...,K\}$, we have
\begin{equation}
H(X_{\mathcal{S}}|Y_{\mathcal{S}^c})\!\geq\! QT \sum\limits_{j=1}^{|{\cal S}|} a^{j,{\cal S}}_{\cal M}  \sum \limits_{\ell = \max\{j,s\}}^{\min\{j+s,|{\cal S}|\}} \frac{{|{\cal S}|-j \choose \ell-j}{j \choose \ell-s}}{{K \choose s}} \cdot \frac{\ell -j}{\ell-1},
\end{equation}
where $a^{j,{\cal S}}_{\cal M}$ is defined in (\ref{eq:countSub}).
$\hfill \square$
\end{claim}

We prove Claim~2 by induction.

a. If $\mathcal{S} = \{k\}$ for any $k \in \{1,\ldots,K\}$, obviously
\begin{align}
H(X_k|Y_{\{1,\ldots,K\}\backslash \{k\}})&\geq 0 =QT a^{1,\{k\}}_{\cal M}  \sum \limits_{\ell = s}^{1} \frac{{0 \choose \ell-1}{1 \choose \ell-s}}{{K \choose s}}.
\end{align}
 
b. Suppose the statement is true for all subsets of size $S_0$. 

For any $\mathcal{S}\subseteq \{1,...,K\}$ of size $|\mathcal{S}|=S_0+1$, and all $k \in \mathcal{S}$, we have as derived in (\ref{eq:twoParts}):
\begin{align}
H(X_{\mathcal{S}}|Y_{\mathcal{S}^c}) \geq& \frac{1}{S_0} \sum_{k\in\mathcal{S}} \Big(H(X_\mathcal{S}|V_{\mathcal{W}_k,:},V_{:,{\mathcal{M}_k}},Y_{\mathcal{S}^c}) \nonumber \\
&+H(V_{\mathcal{W}_k,:}|V_{:,{\mathcal{M}_k}},Y_{\mathcal{S}^c})\Big), \label{eq:int8}
\end{align}
where $Y_{\mathcal{S}^c} =(V_{\mathcal{W}_{\mathcal{S}^c},:},V_{:,\mathcal{M}_{\mathcal{S}^c}})$.
 
The first term on the RHS of (\ref{eq:int8}) is lower bounded by the induction assumption:
\begin{align}
&H(X_\mathcal{S}|V_{\mathcal{W}_k,:},V_{:,{\mathcal{M}_k}},Y_{\mathcal{S}^c})=H(X_{\mathcal{S}\backslash \{k\}}|Y_{(\mathcal{S}\backslash \{k\})^c})\\    
&\geq QT \sum\limits_{j=1}^{S_0} a^{j,{\cal S} \backslash \{k\}}_{\cal M}  \sum \limits_{\ell = \max\{j,s\}}^{\min\{j+s,S_0\}} \frac{{S_0-j \choose \ell-j}{j \choose \ell-s}}{{K \choose s}} \cdot \frac{\ell -j}{\ell-1}. \label{eq:first2}
\end{align} 

The second term on the RHS of (\ref{eq:int8}) can be calculated based on the independence of intermediate values:
\begin{align}
&H(V_{\mathcal{W}_k,:}|V_{:,{\mathcal{M}_k}},Y_{\mathcal{S}^c})\nonumber \\
=&H(V_{\mathcal{W}_k,:}|V_{:,{\mathcal{M}_k}},V_{\mathcal{W}_{\mathcal{S}^c},:},V_{:,\mathcal{M}_{\mathcal{S}^c}})\\
\overset{(a)}{=}&H(V_{\mathcal{W}_k,:}|V_{{\cal W}_k,{\mathcal{M}_k}\cup{\cal M}_{{\cal S}^c}},V_{\mathcal{W}_{\mathcal{S}^c},:})\\
\overset{(b)}{=}&\sum_{q \in {\cal W }_k} H(V_{\{q\},:}|V_{\{q\},{\mathcal{M}_k}\cup{\cal M}_{{\cal S}^c}},V_{\mathcal{W}_{\mathcal{S}^c},:})\\
\overset{(c)}{=}&\frac{Q}{{K \choose s}} {|{\cal S}|-1 \choose s-1}T \sum_{j=0}^{S_0} a^{j,\mathcal{S}\backslash \{k\}}_{\cal M}\\
\geq &\frac{Q}{{K \choose s}} {|{\cal S}|-1 \choose s-1}T\sum_{j=1}^{S_0} a^{j,\mathcal{S}\backslash \{k\}}_{\cal M}, \label{eq:second2}
\end{align}
where (a) and (b) are due to the independence of the intermediate values, and (c) is due to the uniform distribution of the output functions such that each node in ${\cal S}$ calculates $\frac{Q}{{K \choose s}}\cdot {|{\cal S}|-1 \choose s-1}$ output functions computed exclusively by $s$ nodes in ${\cal S}$. 

Thus by (\ref{eq:int8}), (\ref{eq:first2}), and (\ref{eq:second2}), we have
\begin{align}
&H(X_{\mathcal{S}}|Y_{\mathcal{S}^c})\nonumber \\
\geq& \frac{QT}{S_0} \sum_{k\in\mathcal{S}}  \sum_{j=1}^{S_0} a^{j,\mathcal{S}\backslash \{k\}}_{\cal M} \bigg(\sum \limits_{\ell = \max\{j,s\}}^{\min\{j+s,S_0\}} \frac{{S_0-j \choose \ell-j}{j \choose \ell-s}}{{K \choose s}} \cdot \frac{\ell -j}{\ell-1} \nonumber\\
&+ \frac{{S_0 \choose s-1}}{{K \choose s}}\bigg) \\
=& \frac{QT}{S_0} \sum_{j=1}^{S_0} \bigg(\sum \limits_{\ell = \max\{j,s\}}^{\min\{j+s,S_0\}} \frac{{S_0-j \choose \ell-j}{j \choose \ell-s}}{{K \choose s}} \cdot \frac{\ell -j}{\ell-1} + \frac{{S_0 \choose s-1}}{{K \choose s}}\bigg)  \nonumber \\
& \cdot \sum_{k\in\mathcal{S}}  a^{j,\mathcal{S}\backslash \{k\}}_{\cal M} \\
=& QT \cdot \frac{S_0+1-j}{S_0} \sum_{j=1}^{S_0} \bigg(\sum \limits_{\ell = \max\{j,s\}}^{\min\{j+s,S_0\}} \frac{{S_0-j \choose \ell-j}{j \choose \ell-s}}{{K \choose s}} \cdot \frac{\ell -j}{\ell-1} \nonumber\\
&+ \frac{{S_0 \choose s-1}}{{K \choose s}}\bigg)a^{j,\mathcal{S}}_{\cal M} \\
=& QT \sum_{j=1}^{S_0+1} \frac{S_0+1-j}{S_0} \bigg(\sum \limits_{\ell = \max\{j,s\}}^{\min\{j+s,S_0\}} \frac{{S_0-j \choose \ell-j}{j \choose \ell-s}}{{K \choose s}} \cdot \frac{\ell -j}{\ell-1} \nonumber \\
&+ \frac{{S_0 \choose s-1}}{{K \choose s}}\bigg)a^{j,\mathcal{S}}_{\cal M}. \label{eq:int3}
\end{align}

 For each $j \in \{1,\ldots, S_0+1\}$ in (\ref{eq:int3}), we have 
 \begin{align}
 &\frac{S_0+1-j}{S_0} \bigg(\sum \limits_{\ell = \max\{j,s\}}^{\min\{j+s,S_0\}} \frac{{S_0-j \choose \ell-j}{j \choose \ell-s}}{{K \choose s}} \cdot \frac{\ell -j}{\ell-1} + \frac{{S_0 \choose s-1}}{{K \choose s}}\bigg) \nonumber\\
 = & \frac{S_0+1-j}{S_0 {K \choose s}}\Bigg(\sum \limits_{\ell = \max\{j,s\}}^{\min\{j+s,S_0\}} {S_0-j \choose \ell-j}{j \choose \ell-s} \frac{\ell -j}{\ell-1} \nonumber \\
 &+ \sum \limits_{\ell = \max\{j+1,s\}}^{\min\{j+s,S_0+1\}} {S_0-j \choose \ell-j-1} {j \choose \ell-s} \Bigg)\\
=& \frac{S_0+1-j}{S_0 {K \choose s}}\Bigg(\sum \limits_{\ell = \max\{j,s\}}^{\min\{j+s,S_0+1\}} {S_0-j \choose \ell-j}{j \choose \ell-s} \frac{\ell -j}{\ell-1} \nonumber \\
&+ \sum \limits_{\ell = \max\{j,s\}}^{\min\{j+s,S_0+1\}} {S_0-j \choose \ell-j-1} {j \choose \ell-s} \Bigg)\\
=&\frac{1}{{K \choose s}}\sum \limits_{\ell = \max\{j,s\}}^{\min\{j+s,S_0+1\}}{S_0+1-j \choose \ell-j}{j \choose \ell-s} \nonumber \\
&\Bigg(\frac{S_0-\ell+1}{S_0}
\cdot \frac{\ell -j}{\ell-1} + \frac{\ell-j}{S_0}\Bigg)\\
=&\frac{1}{{K \choose s}}\sum \limits_{\ell = \max\{j,s\}}^{\min\{j+s,S_0+1\}}{S_0+1-j \choose \ell-j}{j \choose \ell-s}\frac{\ell-j}{\ell-1}.\label{eq:int4}
 \end{align}	
 
Applying (\ref{eq:int4}) into (\ref{eq:int3}) yields
 \begin{align}
&H(X_{\mathcal{S}}|Y_{\mathcal{S}^c}) \nonumber \\
\geq& QT \sum\limits_{j=1}^{S_0+1} a^{j,{\cal S}}_{\cal M}  \sum \limits_{\ell = \max\{j,s\}}^{\min\{j+s,S_0+1\}} \frac{{S_0+1-j \choose \ell-j}{j \choose \ell-s}}{{K \choose s}} \cdot \frac{\ell -j}{\ell-1} \\ 
& =QT \sum\limits_{j=1}^{|{\cal S}|} a^{j,{\cal S}}_{\cal M}  \sum \limits_{\ell = \max\{j,s\}}^{\min\{j+s,|{\cal S}|\}} \frac{{|{\cal S}|-j \choose \ell-j}{j \choose \ell-s}}{{K \choose s}} \cdot \frac{\ell -j}{\ell-1}.\label{eq:int5}
\end{align}
	
Since (\ref{eq:int5}) holds for all subsets ${\cal S}$ of size $|{\cal S}| = S_0+1$, we have proven Claim~2.

Then by Claim~2, let $\mathcal{S}=\{1,...,K\}$ be the set of all $K$ nodes, 
\begin{align}
L^*_\mathcal{M}(s)&\geq \frac{H(X_{\mathcal{S}}|Y_{\mathcal{S}^c})}{QNT} \nonumber \\
&\geq \sum_{j=1}^{K} \frac{a^{j}_{\cal M}}{N} \sum \limits_{\ell = \max\{j,s\}}^{\min\{j+s,K\}} \frac{{K-j \choose \ell-j}{j \choose \ell-s}}{{K \choose s}} \cdot \frac{\ell -j}{\ell-1}. 
\end{align}

This completes the proof of Lemma~2. $\hfill \blacksquare$

\section{Implementation and Empirical Evaluation of Coded Distributed Computing}\label{sec:implementation}

In this section, we demonstrate the impact of the proposed Coded Distributed Computing (CDC) scheme on balancing the time spent on task execution and the time spent on data movement, in order to speed up practical distributed computing applications. In particular, let us consider a MapReduce-type application for which the total execution time is roughly composed of the time spent executing the Map tasks, denoted by $T_{\textup{map}}$, the time spent shuffling intermediate values, denoted by $T_{\textup{shuffle}}$, and the time spent executing the Reduce tasks, denoted by $T_{\textup{reduce}}$, i.e.,
\begin{equation}
\label{eq:totalMR}
T_{\textup{total, MR}} \approx T_{\textup{map}}+ T_{\textup{shuffle}}+ T_{\textup{reduce}}.
\end{equation}

Using CDC, we can leverage $r\times$ more computations in the Map phase, in order to reduce the communication load by the same multiplicative factor. Hence, ignoring the coding overheads, CDC promises an approximate total execution time of
\begin{align}\label{eq:total}
T_{\textup{total, CDC}} \approx rT_{\textup{map}}+ \tfrac{1}{r}T_{\textup{shuffle}}+ T_{\textup{reduce}}.
\end{align}

To minimize the above execution time, one would choose $r^* =\left\lfloor \sqrt{\tfrac{T_{\textup{shuffle}}}{T_{\textup{map}}}} \right\rfloor \textup{ or } \left\lceil \sqrt{\tfrac{T_{\textup{shuffle}}}{T_{\textup{map}}}} \right\rceil$, resulting in the minimum execution time of
\begin{equation}\label{eq:totalCMR}
T^*_{\textup{total, CDC}} \approx 2\sqrt{T_{\textup{shuffle}}T_{\textup{map}}}+T_{\textup{reduce}}.
\end{equation}
For example, in an application that $T_{\textup{shuffle}}$ is $10\times$ - $100\times$ larger than $T_{\textup{map}}+T_{\textup{reduce}}$,  by comparing from (\ref{eq:totalMR}) and (\ref{eq:totalCMR}), we note that CDC can reduce the execution time by approximately $1.5 \times$ - $5 \times$. 

In the rest of this section, we empirically demonstrate the performance gain of applying CDC to \texttt{TeraSort}~\cite{TSpackage}, which is a commonly used Hadoop benchmark for distributed sorting terabytes of data~\cite{Terasort}. In particular, we first incorporate the coding ideas in CDC into \texttt{TeraSort} to develop a novel coded distributed sorting algorithm, named \texttt{CodedTeraSort}, which imposes \emph{structured} redundancy in the input data, in order to enable in-network coding opportunities that overcome the data shuffling bottleneck of \texttt{TeraSort}. Then, we evaluate the performance of \texttt{CodedTeraSort} on Amazon EC2 clusters, and observe a $1.97\times$ - $3.39\times$ speedup, compared with \texttt{TeraSort}, for typical settings of interest.

\subsection{TeraSort}\label{sec:TeraSort}
\texttt{TeraSort}~\cite{Terasort} is a conventional algorithm for distributed sorting of a large amount of data. The input data that is to be sorted is in the format of key-value (KV) pairs, meaning that each input KV pair consists of a key and a value. For example, the domain of the keys can be 10-byte integers, and the domain of the values can be arbitrary strings. \texttt{TeraSort} sorts the input data according to their keys, e.g., sorting integers. 

\subsubsection{Algorithm Description} 
\begin{figure}[htbp]
  \centering
  \includegraphics[width=0.48\textwidth]{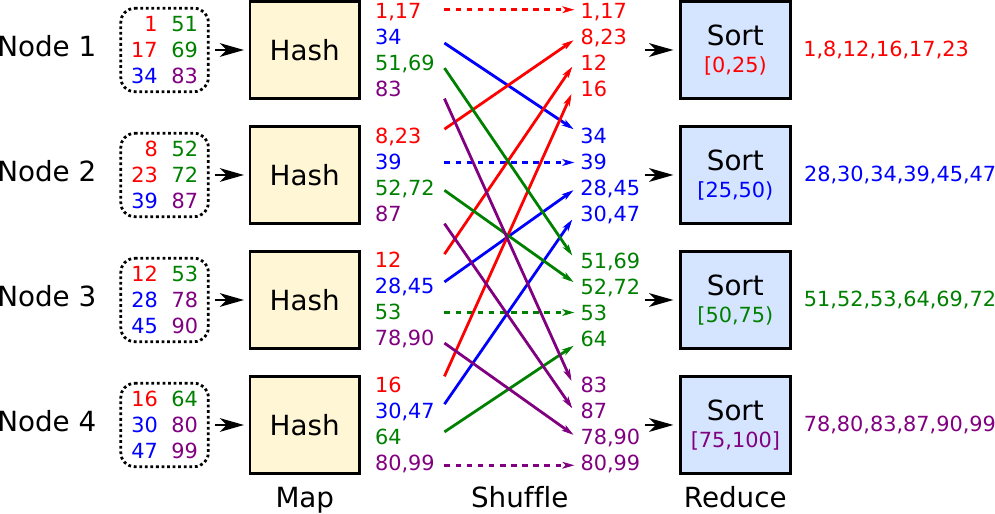}
  \caption{Illustration of \texttt{TeraSort} algorithm with $K = 4$ nodes and key domain partitions $[0,25), [25,50), [50,75), [75,100]$.  A dotted box represents an input file.  An input file is hashed into 4 groups of KV pairs, one for each partition. For each of the 4 partitions, the KV pairs belonging to that partition computed on all 4 nodes are fetched to a corresponding node, which sorts all KV pairs in that partition locally.}
  \label{fig:terasort}
\end{figure}

Let us consider implementing \texttt{TeraSort} over $K$ distributed computing nodes, which consists of 5 stages: File Placement, Key Domain Partitioning, Map Phase, Shuffle Phase, and Reduce Phase.  In File Placement, all input KV pairs are split into $K$ disjoint files, and each file is placed on one of the $K$ nodes.  In Key Domain Partitioning, the domain of the keys is split into $K$ partitions, and each node will be responsible for sorting the KV pairs whose keys fall into one of the partitions. In Map Phase, each node hashes each KV pair in its locally stored file into one of the $K$ partitions, according to its key. In Shuffle Phase, the KV pairs in the same partition are transferred to the node that is responsible for sorting that partition. In Reduce Stage, each node locally sorts KV pairs belonging to its assigned partition. We illustrate the \texttt{TeraSort} algorithm using a simple example shown in Fig. \ref{fig:terasort}.  

\subsubsection{Performance Evaluation} 
To understand the performance of \texttt{TeraSort}, we performed an experiment on Amazon EC2 to sort 12GB of data by running \texttt{TeraSort} on 16 instances.\footnote{We note that EC2 uses virtual machines, and each instance may not be hosted by a dedicated physical machine.} The breakdown of the total execution time is shown in Table \ref{tlb:exampleTeraSort}.

\begin{table}[htbp]
    \centering  
    \caption{Performance of \texttt{TeraSort} sorting $12$GB data with $K=16$ instances and $100$ Mbps network speed} 
    \label{tlb:exampleTeraSort}
    \begin{tabular}{|c|c|c|c|c|c|}\hline
    Map    & Pack   & Shuffle  & Unpack  & Reduce  & Total \\
    (sec.) & (sec.) & (sec.)   & (sec.)  & (sec.)  & (sec.) \\\hline
    1.86   & 2.35   & 945.72   & 0.85    & 10.47   & 961.25 \\\hline
    \end{tabular}
\end{table}

We observe from Table \ref{tlb:exampleTeraSort} that for a conventional \texttt{TeraSort} execution, 98.4\% of the total execution time was spent in data shuffling, which is $508.5 \times$ of the time spent in the Map phase. Given the fact that data shuffling dominates the job execution time, the principle of optimally trading computation for communication of the proposed CDC scheme can be applied to significantly improve the performance of \texttt{TeraSort}. For example, when executing the same sorting job using a coded version of \texttt{TeraSort} with a computation load of $r = 10$, according to (\ref{eq:total}), we could theoretically save the total execution time by approximately $8 \times$. This motivates us to develop a novel coded distributed sorting algorithm, named \texttt{CodedTeraSort}, which is briefly described in the next sub-section.

\subsection{Coded TeraSort}\label{sec:codedTeraSort}
We develop the \texttt{CodedTeraSort} algorithm by applying the proposed CDC scheme for the case of $s=1$ (see Example 1 in Section~\ref{sec:example} for an illustration) to the above described \texttt{TeraSort} algorithm. \texttt{CodedTeraSort} exploits redundant computations on the input files in the Map phase, creating in-network coding opportunities to significantly slash the load of data shuffling. In particular, the execution of \texttt{CodedTeraSort} consists of following $6$ stages of operations. Here we give high-lever descriptions of these operations, and we refer the interested readers to~\cite{li2017coded} for more detailed descriptions.
\begin{enumerate}
    \item \emph{Structured Redundant File Placement}. The entire input KV pairs are split into many small files, each of which is repeatedly placed on $1 \leq r \leq K$ nodes (i.e., a computation load of $r$), according to the particular pattern specified by the CDC scheme. 
    \item \emph{Map}. Each node applies the hashing operation as in \texttt{TeraSort} on each of its assigned files.
    \item \emph{Encoding to Create Coded Packets}. Each node generates coded multicast packets from local results computed in Map phase, according to the encoding process of the CDC scheme.
    \item \emph{Multicast Shuffling}. Each node multicasts each of its generated coded packet to a specific set of $r$ other nodes.
    \item \emph{Decoding}. Each node locally decodes the required KV pairs from the received coded packets.
    \item \emph{Reduce}. Each node locally sorts the KV pairs within its assigned partition as in the Reduce phase of \texttt{TeraSort}.
\end{enumerate}

\subsection{Empirical Evaluations}\label{sec:evaluation}

We imperially demonstrate the performance gain of \texttt{CodedTeraSort} through experiments on Amazon EC2 clusters.  In this sub-section, we first present some choices we have made for the implementation. Then, we discuss the experiment results.

\subsubsection{Implementation Choices}

We first describe the following common implementation choices that we have made for both \texttt{TeraSort} and \texttt{CodedTeraSort} algorithms.

\emph{Data Format:} All input KV pairs are generated from \texttt{TeraGen}~\cite{TSpackage} in the standard Hadoop package.  Each input KV pair consists of a $10$-byte key and a $90$-byte value. A key is a $10$-byte unsigned integer, and the value is an arbitrary string of $90$ bytes. The KV pairs are sorted based on their keys, using the standard integer ordering.

\emph{Library:} We implement both \texttt{TeraSort} and \texttt{CodedTeraSort} algorithms in \texttt{C++}, and use Open MPI library \cite{openMPI} for communications between EC2 instances.

\emph{System Architecture:} We employ a system architecture that consists of a coordinator node and $K$ worker nodes, for some $K \in \mathbb{N}$. Each node is run as an EC2 instance. The coordinator node is responsible for creating the key partitions and placing the input files on the local disks of the worker nodes. The worker nodes are responsible for distributedly executing the stages of the sorting algorithms. 

\begin{figure}[htbp]
  \centering
  \includegraphics[width=0.3\textwidth]{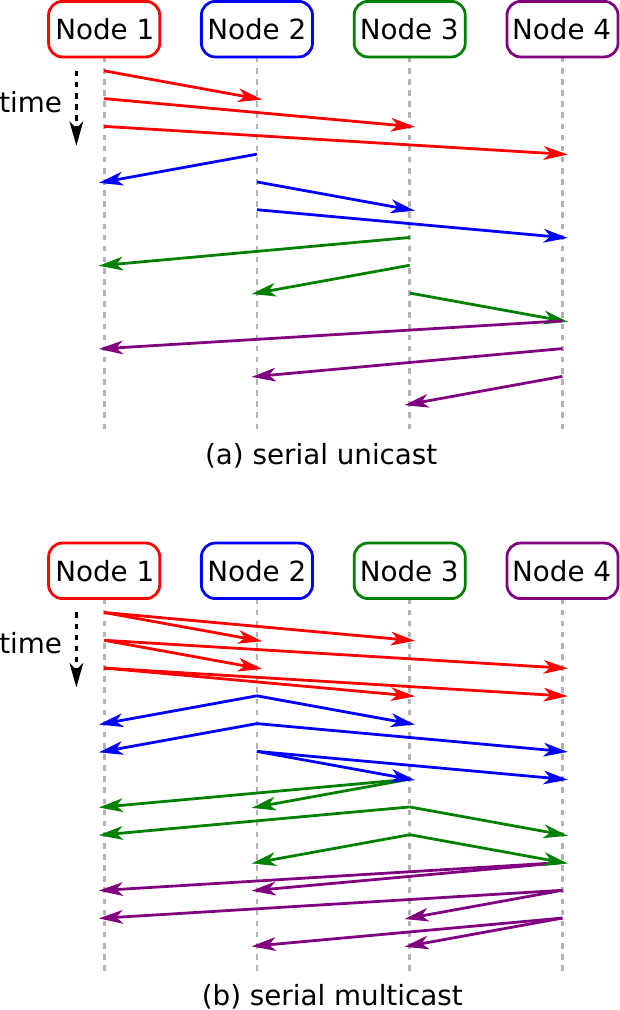}
  \caption{(a) Serial unicast in the Shuffle phase of \texttt{TeraSort}; a solid arrow represents a unicast. (b) Serial multicast in the Multicast Shuffle phase of \texttt{CodedTeraSort}; a group of solid arrows starting at the same node represents a multicast.}
  \label{fig:serial}
\end{figure}

In the \texttt{TeraSort} implementation, each node sequentially steps through Map, Pack, Shuffle, Unpack, and Reduce stages.
The Pack stage serializes each intermediate value to a continuous memory array to ensure that a single TCP flow is created for each intermediate value (which may contain multiple KV pairs) when \texttt{MPI\_Send} is called\footnote{Creating a TCP flow per KV pair leads to inefficiency from overhead and convergence issue.}.  The Unpack stage deserializes the received data to a list of KV pairs.  In the Shuffle stage, intermediate values are unicast serially, meaning that there is only one sender node and one receiver node at any time instance.  Specifically, as illustrated in Fig. \ref{fig:serial}(a), Node $1$ starts to unicast to Nodes 2, 3, and 4  back-to-back.  After Node $1$ finishes, Node $2$ unicasts back-to-back to Nodes 1, 3, and 4.  This continues until Node 4 finishes. 

In the \texttt{CodedTeraSort} implementation, each node sequentially steps through CodeGen, Map, Encode, Multicast Shuffling, Decode, and Reduce stages. In the CodeGen (or code generation) stage, firstly, each node generates all file indices, as subsets of $r$ nodes. Then each node uses \texttt{MPI\_Comm\_split} to initialize $\binom{K}{r+1}$ multicast groups each containing $r+1$ nodes on Open~MPI, such that multicast communications will be performed within each of these groups.  The serialization and deserialization are implemented respectively in the Encode and the Decode stages.  In Multicast Shuffling, $\texttt{MPI\_Bcast}$ is called to multicast a coded packet in a serial manner, so only one node multicasts one of its encoded packets at any time instance.  Specifically, as illustrated in Fig. \ref{fig:serial}(b), Node 1 multicasts to the other 2 nodes in each multicast group Node 1 is in. For example, Node 1 first multicasts to Node 2 and 3 in the multicast group $\{1,2,3\}$.  After Node $1$ finishes, Node $2$ starts multicasting in the same manner. This process continues until Node $4$ finishes.

\subsubsection{Experiment Results}

\begin{table*}[!t]
\centering
\caption{Sorting $12$ GB data with $K = 16$ worker instances and 100 Mbps network speed}
  \label{tlb:K16}
  \begin{tabular}{|c|c|c|c|c|c|c|c|c|}
    \hline
             & CodeGen & Map & Pack/Encode & Shuffle & Unpack/Decode & Reduce & Total Time & Speedup \\
             & (sec.)   & (sec.) & (sec.) & (sec.) & (sec.) & (sec.) & (sec.) & \\\hline
    \texttt{TeraSort}:           &  --   & 1.86  & 2.35 & 945.72  & 0.85 & 10.47 & 961.25 & \\
    \texttt{CodedTeraSort}: $r=3$& 6.06	 & 6.03  & 5.79 & 412.22  & 2.41 & 13.05 & 445.56 & 2.16$\times$ \\
    \texttt{CodedTeraSort}: $r=5$& 23.47 & 10.84 & 8.10 & 222.83  & 3.69 & 14.40 & 283.33 & 3.39$\times$ \\\hline
  \end{tabular}
\end{table*}

\begin{table*}[!t]
\centering
\caption{Sorting $12$ GB data with $K = 20$ worker instances and 100 Mbps network speed}
  \label{tlb:K20}
  \begin{tabular}{|c|c|c|c|c|c|c|c|c|}
    \hline
             & CodeGen & Map & Pack/Encode & Shuffle & Unpack/Decode & Reduce & Total Time & Speedup \\
             & (sec.)   & (sec.) & (sec.) & (sec.) & (sec.) & (sec.) & (sec.) & \\\hline
    \texttt{TeraSort}:           &  --  & 1.47 & 2.00 & 960.07 & 0.62 & 8.29 & 972.45 & \\
    \texttt{CodedTeraSort}: $r=3$& 19.32 & 4.68 & 4.89 & 453.37 & 1.87 & 9.73 & 493.86 & 1.97$\times$ \\
    \texttt{CodedTeraSort}: $r=5$& 140.91 & 8.59 & 7.51 & 269.42 & 3.70 & 10.97 & 441.10 & 2.20$\times$ \\\hline
  \end{tabular}
\end{table*}

We evaluate the run-time performance of \texttt{TeraSort} and \texttt{CodedTeraSort}, for different combinations of the number of workers $K$ and the computation load $1 \leq r \leq K$. All experiments are repeated $5$ times, and the average values are recorded. 

In Table~\ref{tlb:K16} and Table~\ref{tlb:K20}, we list the breakdowns of the average execution times to sort 12 GB of input data using $K=16$ workers and $K=20$ workers respectively. Here we limit the incoming and outgoing traffic rates of each instance to $100$ Mbps. This is to alleviate the effects of the bursty behaviors of the transmission rates in the beginning of some TCP sessions, given the particular size of the data to be sorted. We observe an overall $1.97\times$ - $3.39\times$ speedup of \texttt{CodedTeraSort} as compared with \texttt{TeraSort}.  From the experiment results we make the following observations:
\begin{itemize}
    \item For \texttt{CodedTeraSort}, the time spent in the CodeGen stage is proportional to 
    $\binom{K}{r+1}$, which is the number of multicast groups. 
    \item The Map time of \texttt{CodedTeraSort} is approximately $r$ times higher than that of \texttt{TeraSort}. This is because that each node hashes $r$ times more KV pairs than that in \texttt{TeraSort}.  Specifically, the ratios of the \texttt{CodedTeraSort}'s Map time to the \texttt{TeraSort}'s Map time from Table \ref{tlb:K16} are $6.03/1.86 \approx 3.2$ and $10.84/1.86 \approx 5.8$, and from Table \ref{tlb:K20} are $4.68/1.47 \approx 3.2$ and $8.59/1.47 \approx 5.8$.
    \item While \texttt{CodedTeraSort} theoretically promises a factor of more than $r \times$ reduction in shuffling time, the actual gains observed in the experiments are slightly less than $r$. For example, for the experiment with $K=16$ nodes and $r=3$, as shown in Table \ref{tlb:K16}, the speedup of the Shuffle stage is $945.72/412.22 \approx 2.3 < 3$. This phenomenon is caused by the following two factors. 1) Open MPI's multicast API ($\texttt{MPI\_Bcast}$) has an inherent overhead per a multicast group, for instance, a multicast tree is constructed before multicasting to a set of nodes.  2) Using the $\texttt{MPI\_Bcast}$ API, the time of multicasting a packet to $r$ nodes is higher than that of unicasting the same packet to a single node. In fact,
    as measured in~\cite{lee2015speeding}, the multicasting time increases logarithmically with $r$.
\end{itemize}

Further, we observe the following trends from both tables:

\emph{The impact of computation load $r$:} As $r$ increases, the shuffling time reduces by approximately $r$ times. However, the  Map execution time increases linearly with $r$, and more importantly the CodeGen time increases exponentially with $r$ as $\binom{K}{r+1}$. Hence, for small values of $r$ ($r < 6$) we observe overall reduction in execution time, and the speedup increases. However, as we further increase $r$, the CodeGen time will dominate the execution time, and the speedup decreases. Hence, in our evaluations, we have limited $r$ to be at most $5$.\footnote{The redundancy parameter $r$ is also limited by the total storage available at the nodes. Since for a choice of redundancy parameter $r$, each piece of input KV pairs should be stored at $r$ nodes, we can not increase $r$ beyond $\frac{\text{total available storage at the worker nodes}}{\text{input size}}.$}

\emph{The impact of worker number $K$:} As $K$ increases, the speedup decreases. This is due to the following two reasons. 1) The number of multicast groups, i.e., $\binom{K}{r+1}$, grows exponentially with $K$, resulting in a longer execution time of the CodeGen process. 2) When more nodes participate in the computation, for a fixed $r$, less amount of KV pairs are hashed at each node locally in the Map phase, resulting in less locally available intermediate values and a higher communication load. Hence, given more worker nodes, one would preferably use larger computation load to achieve a better run-time performance.
    
\section{Concluding Remarks and Future Directions}
We introduced a scalable distributed computing framework motivated by MapReduce, which is suited for arbitrary types of output functions. We formulated and exactly characterized an information-theoretic tradeoff between computation load and communication load within this framework. In particular, we proposed Coded Distributed Computing (CDC), a coded scheme that reduces the communication load by a factor that can grow with the network size, illustrating the role of coding in speeding up distributed computing jobs. We also proved a tight information-theoretic lower bound on the minimum communication load, using any data shuffling scheme, which exactly matches the communication load achieved by CDC. This result reveals a fundamental relationship between computation and communication in distributed computing--the two are inversely proportional to each other. Moreover, we applied the proposed CDC scheme to the conventional \texttt{TeraSort} algorithm to develop a novel distributed sorting algorithm, named \texttt{CodedTeraSort}, and empirically demonstrated the performance gain of \texttt{CodedTeraSort} through extensive experiments on Amazon EC2 clusters.

Finally, we discuss some follow-up research directions of this work. 

{\bf Heterogeneous Networks with Asymmetric Tasks.}
It is common to have computing nodes with heterogeneous storage, processing and communication capacities within computer clusters (e.g., Amazon EC2 clusters composed of heterogeneous computing instances). In addition, processing different parts of the dataset can generate intermediate results with different sizes (e.g., performing data analytics on highly-clustered graphs). For computing over heterogeneous nodes, one solution is to break the more powerful nodes into multiple  smaller virtual nodes that have homogeneous capability, and then apply the proposed CDC scheme for the homogeneous setting. When intermediate results have different sizes, the proposed coding scheme still applies, but the coding operations are not symmetric as in the case of homogeneous intermediate results (e.g., one may now need to compute the XOR of two data segments with different sizes). Alternatively, we can employ a low-complexity greedy approach, in which we assign the Map tasks to maximize the number of multicasting opportunities that simultaneously deliver useful information to the largest possible number of nodes. Some preliminary studies along this direction have been conducted to obtain the solutions for some special cases (see, e.g.,~\cite{reisizadeh2017coded,kiamari2017heterogeneous}). Nevertheless, systematically characterizing the optimal resource allocation strategies and coding schemes for general heterogeneous networks with asymmetric tasks remains an interesting open problem.

{\bf Straggling/Failing Computing Nodes.} 
Other than the communication bottleneck, the effect of straggling servers also severely degrades the run-time performance of distributed computing applications (see e.g.,~\cite{zaharia2008improving}). Recently in~\cite{lee2015speeding}, Maximum-Distance-Separable (MDS) codes were utilized to encode linear computation tasks, providing robustness to a certain number of stragglers. Following the results in~\cite{lee2015speeding}, coded computing strategies have been proposed to efficiently deal with the stragglers for various computation tasks and network settings (see, e.g.,~\cite{dutta2016short,lee2017high,yu2017polynomial,pmlr-v70-tandon17a}). In~\cite{LMA16_unify}, we have superimposed the proposed CDC scheme on top of the MDS codes, developing a unified coding framework for distributed computing with straggling servers. This framework achieves a flexible tradeoff between computation latency in the Map phase and communication load in the Shuffle phase, which has the CDC scheme (or minimum bandwidth code) and the MDS code (or minimum latency code) as the two end points. Nevertheless, designing resource allocation strategies and coding techniques to optimize the run-time performance over distributed computing clusters with stragglers is a challenging open problem.

{\bf Multi-Stage Computation Tasks.} 
Unlike simple computation tasks like Grep, Join and Sort, many distributed computing applications contain multiples stages of MapReduce computations. Examples of these applications include machine learning algorithms~\cite{chu2007map}, SQL queries for databases~\cite{isard2007dryad,abouzeid2009hadoopdb}, and scientific analytics~\cite{ekanayake2009dryadlinq}. One can express the computation logic of a multi-stage application as a directed acyclic graph (DAG)~\cite{saha2015apache}, in which each vertex represents a logical step of data transformation, and each edge represents the dataflow across processing vertices. In order to speed up multi-stage computation tasks using codes, while one straightforward approach is to apply the proposed CDC scheme for the cascaded distributed computing framework (see Theorem~2) to compute each stage locally, we expect to achieve a higher reduction in bandwidth consumption and response time by globally designing codes for the entire task graph and accounting for interactions between consecutive stages. A preliminary exploration along this direction was recently presented in~\cite{LMA-Allerton16}.

{\bf Multi-Layer Networks and Structured Topology.} So far we have only considered a single-layer topology of the distributed computing nodes, in which each node can multicast to an arbitrary number of other nodes at the same cost as unicasting to a single node. However, in practical data center networks, nodes can be connected through multiple switches at different layers with different capacities, forming a hierarchical multi-root tree topology (e.g., fat-tree topology~\cite{al2008scalable}). In this case, we need to generalize our communication model to include more structured topologies, and develop coded shuffling strategies that account for (1) path lengths of shuffled data (2) congestion at links higher up in the topology; and (3) different link capacities and multicast-costs at different layers of network topology. We have made preliminary progress in~\cite{li2016scalable} for a star topology (motivated by wireless edge computing), where nodes are connected via only one access point (or switch layer).

{\bf Joint Storage and Computation Optimization.}
We have so far assumed that we can design the placement of the input files to create coding opportunities during the computation process. However, in practical file storage systems, data blocks are often stored without prior knowledge about the computations that will be performed on them, and moving the data across the nodes before the computation is often too costly. In this case, even without the capability of designing the data placement as exactly specified by the CDC scheme, one can still take advantage of the inherent data redundancy (e.g., GFS~\cite{ghemawat2003google} and HDFS~\cite{hdfs} by default place replicas of each data block on 3 distributed nodes) to create coded multicast opportunities, significantly reducing the communication load. 

\begin{figure}[htbp]
   \centering
   \includegraphics[width=0.4\textwidth]{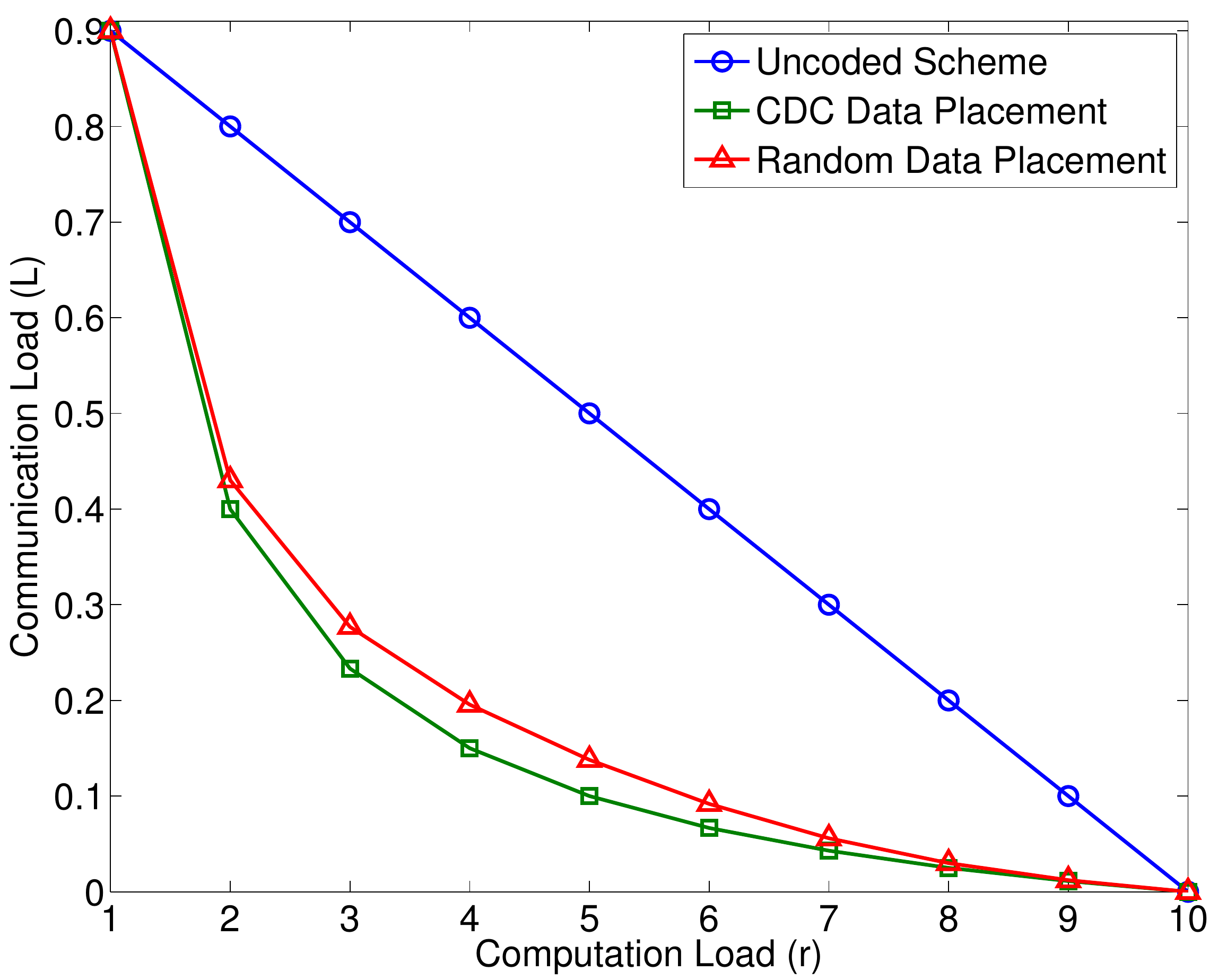}
   \caption{Comparison of the average communication load by placing and mapping every input file randomly at $r$ out of $K=10$ nodes with the communication load achieved by placing the files specified by the CDC scheme. Here we compute $Q=10$ output functions from 2520 input files using $K=10$ distributed computing nodes.}
   \label{fig:random}

\end{figure}

We plot in Fig.~\ref{fig:random} the average communication load achieved by a coded shuffling scheme similar to the one presented in Section~\ref{sec:shuffle} (with the modification that each node zero-pads its associated data segments to the length of the longest one before coding), when each input file is placed and mapped at $r$ out of $K$ nodes chosen \emph{uniformly at random}, and compare it with the communication load achieved by CDC where the input files are placed based on the Map phase design in Section~\ref{sec:map}. As demonstrated in Fig.~\ref{fig:random}, without requiring the files to be placed as exactly described by the CDC scheme, one can still exploit the data redundancy to achieve a communication load that is superlinear with respect to the computation load. Therefore, the coded data shuffling scheme of CDC can effectively reduce the communication loads of computation jobs on general data storage systems. This behavior that a random data placement achieves close-to-optimum performance has also been reported in~\cite{li2016scalable} for a decentralized wireless distributed computing platform, and in~\cite{maddah2013decentralized} for a decentralized caching system.

{\bf Coded Edge/Fog Computing.}
In the emerging mobile Edge/Fog computing paradigm (see, e.g.,~\cite{bonomi2012fog,CZ}), abundant computation resources scattered across the network edge (e.g., smartphones, tablets and smart cars) are harvested to perform data-intensive computations collaboratively. In this scenario, coding opportunities are widely available by injecting redundant storage and computations into the edge network. We envision codes to play a transformational role in Edge/Fog computing for leveraging such redundancy to substantially reduce the bandwidth consumption and the latency of computing. For an edge computing scenario where the mobile users upload the tasks to the edge nodes, and retrieve the computed results from the edge nodes, we have designed coded computing architectures in~\cite{li2017communication,FWC17}, in which coded computations that are aware of the underlying physical-layer communication are performed at the edge nodes, achieving the minimum load of computation and the maximum spectral efficiency simultaneously. In~\cite{li2016scalable}, we have formulated a wireless distributed computing framework, in which a cluster of mobile users collaborate via an access point to simultaneously meet their computational needs. For this wireless computing platform, we exploited the coding techniques of CDC to achieve a scalable design such that the platform can accommodate an unlimited number of mobile users with a constant amount of bandwidth consumption. Also in a recent magazine paper~\cite{li2017codingfog}, we have demonstrated the opportunities of utilizing coding to improve the performance of Edge/Fog computing applications (e.g., navigation services and recommendation systems).

\bibliographystyle{IEEEtran}
\bibliography{ref}

\end{document}